\documentclass[a4paper, 11pt]{article}

\usepackage{jcappub}
\usepackage[top=30truemm,bottom=40truemm,left=20truemm,right=20truemm]{geometry}

\usepackage{eucal}
\usepackage{mathrsfs}
\usepackage{mathtools, amsmath, amssymb, bm, bbold}

\usepackage[T1]{fontenc}
\usepackage{newtxtext,newtxmath}

\allowdisplaybreaks[4]

\title{\sc Primordial black holes and \\smooth coarse-graining in excursion set theory}

\author[\boldmath \spadesuit]{Daiki Saito}
\author[\boldmath \clubsuit]{and Koki Tokeshi}

\affiliation[\boldmath \clubsuit]{\it Department of Physics, Kyoto University, Kyoto 606-8502, Japan}
\affiliation[\boldmath \spadesuit]{\it Institute for Cosmic Ray Research (ICRR), The University of Tokyo, Chiba 277-8582, Japan}

\emailAdd{saito@tap.scphys.kyoto-u.ac.jp}
\emailAdd{tokeshi@icrr.u-tokyo.ac.jp}

\abstract{
	The excursion-set formalism enables us to infer the mass distribution of collapsed objects, such as primordial black holes (PBHs), by the language of stochastic processes.  
	Within the framework, this article investigates how a smooth coarse-graining procedure affects the resulting PBH mass function. 
    As a demonstrative example, we employ a Gaussian window function, for which the stochastic noise becomes fully correlated across scales. 
    It is found that these correlated noises result in a mass function of PBHs, whose maximum and its neighbourhood are predominantly determined by the probability that the density contrast exceeds a given threshold at each mass scale. 
    Our results clarify the role of noise correlations induced by smooth coarse-graining and highlight their importance in predicting the abundance of PBHs. 
}

\begin{document}

\maketitle
\flushbottom

\section{Introduction}
\label{sec:intro}

Primordial black hole (PBH) is a hypothetical black hole (BH) that may be formed in the very early universe~\cite{Zeldovich:1967lct, Hawking:1971ei, Hawking:1974rv, Carr:1974nx, Carr:1975qj, Chapline:1975ojl} (see Refs.~\cite{Escriva:2022duf, Yoo:2022mzl, Harada:2024jxl, Carr:2025kdk} for recent reviews). 
While various formation mechanisms have been proposed so far, most of them rely on an enhanced density contrast that, if it exceeds a certain threshold, leads to direct collapse especially during the radiation-dominated era. 
Such an overdense region could, for instance, originate from the scalar field that drives cosmic inflation~\cite{Starobinsky:1980te, Sato:1980yn, Guth:1980zm, Linde:1981mu, Albrecht:1982wi, Linde:1983gd}. 
When there exists an inflexion point or an exactly flat region on a much smaller scale than those probed by large-scale observations~\cite{Planck:2018jri, Planck:2019kim}, the primordial fluctuation gets amplified so that PBHs could be formed. 
In such cases, various cosmological conundrums, such as the origin of the merger of BHs~\cite{Bird:2016dcv, Clesse:2016vqa, Sasaki:2016jop} found by the first and direct detection of gravitational waves~\cite{LIGOScientific:2016aoc}, could be explained by PBHs depending on the non-trivial structure of the potential on small scales, see \textit{e.g.}~Refs.~\cite{Starobinsky:1992ts, Ivanov:1994pa, Garcia-Bellido:1996mdl, Ivanov:1997ia, Yokoyama:1998pt, Kanazawa:2000ea, Inoue:2001zt, Kinney:2005vj, Martin:2012pe, Namjoo:2012aa, Garcia-Bellido:2017mdw, Germani:2017bcs, Motohashi:2017kbs,Pi:2017gih}. 

One of the most important quantities to be estimated is the mass function of PBHs, since it can be constrained by observations~\cite{Carr:2020gox}. 
Several formulations and methods have been used to derive the mass function of PBHs, including Carr's formula~\cite{Carr:1975qj}, Press--Schechter formula~\cite{Press:1973iz}, excursion-set method~\cite{Bond:1990iw} (see Ref.~\cite{Zentner:2006vw} for review), and the statistics of random Gaussian fields~\cite{Bardeen:1985tr}. 
Those formulations would not result in the same mass function in general, sometimes giving rise to discrepancies in orders. 
For instance, there is a factor-two difference between the Press--Schechter and excursion-set mass functions, which is 
related to the so-called cloud-in-cloud problem discussed in the literature. 
Though the uncertainty of the choice of the formulation itself can affect the predicted abundance of PBHs~\cite{Young:2014ana}, even after one fixes the method to evaluate the abundance there remain uncertainties such as the choice of the window function~\cite{Ando:2018qdb, Tokeshi:2020tjq} (see also Refs.~\cite{Young:2019osy, Pi:2024ert}). 
The window function smooths the density contrast over a characteristic scale, and is needed to relate the mass of a PBH and the scale of the horizon re-entry of a mode at which a PBH with the corresponding mass is assumed to have formed. 
However, it seems that there is neither consensus nor a guiding principle for the adequate choice of the window function. 

The excursion-set method identifies the mass of some virialised object as the first-crossing (or first-passage) time, in the language of stochastic processes~\cite{Chandrasekhar:1943ws}. 
When it is applied to the context of halo formation, aside from PBHs~\cite{Green:1999xm, MoradinezhadDizgah:2019wjf, DeLuca:2020ioi, Auclair:2020csm, Erfani:2021rmw, Auclair:2024jwj, Kameli:2025qzp, Kushwaha:2025zpz}, it is well known that the Heaviside window function in Fourier space significantly simplifies mathematical treatment, as well as resolving the factor-two issue in the Press--Schechter formalism. 
In that case, the coarse-grained density contrast is subjected to the inflow of the modes of larger wavelength as one decreases the coarse-graining scale, giving the random and uncorrelated increment or decrement. 
The situation is altered when a different window function is employed, since in such cases the variation of the coarse-graining scale corresponds to the variation of the weight of the integrated modes, instead of the one-direction inflow of the modes. 
In other words, smoothing the density contrast by a window function other than the Fourier-space Heaviside function inevitably introduces the correlation between scales, described by correlated or \textit{coloured} noises. 

However, the situation becomes different when the excursion-set method is applied to the formation of PBHs, as was very recently pointed out in Ref.~\cite{Kushwaha:2025zpz}. 
The difference mainly comes from the fact that a PBH could be formed when the enhanced density contrast with the relevant scale re-enters the horizon, introducing the correlation between scales even when one employs the Fourier-space Heaviside window function, in addition to the uncorrelated noise often focussed on in the literature. 
Besides, when a smooth window function is employed, scales are completely correlated and the underlying probabilities that constitute the mass function could be considerably altered. 
Such a situation and the associated consequences are the main interest of the present article.  
In particular, the excursion-set ``evolution'' as well as the resultant mass function of PBHs under the Gaussian window function are demonstrated. 
The Gaussian window function, along with the Heaviside window function in Fourier or real space, is one of the most widely used filters in the context of the formation of either halos or PBHs. 
It enables one to derive the covariance matrix of the noise analytically, so that the numerical generation of the correlated, or coloured, noises as well as the numerical reconstruction of the mass function remain computationally inexpensive. 

Correlation between scales in the context of PBH formation results in several interesting implications. 
Our findings include that, for the smooth window function, there is no contribution from uncoloured noise and that PBHs would be formed on scales smaller than the fiducial scale of the power spectrum. 
The mass function around its maximum is, as a consequence, determined mainly by the threshold-piercing probability, which implies that the fudge factor is given approximately by unity. 
This is in contrast to the case in which the Fourier-space Heaviside window function is used, where both the threshold-piercing probability and its ``mirror image'' nearly equally contribute to the relevant part of the mass function. 
While in any case there is no longer the Markov property in the stochastic process in the context of the formation of PBHs, such a difference originates from the choice of the window function. 
It can equivalently be attributed to the existence or non-existence of the contribution from uncoloured noises. 

The organisation of the present article is as follows. 
In Section~\ref{sec:stexs}, the standard excursion-set formalism is reviewed, including a short discussion of how the correlated noises emerge through the coarse-graining process. 
The former half of Section~\ref{sec:pbexs} reviews the excursion-set formalism in the context of PBH formation, where the necessary quantities are derived without specifying the window function or the power spectrum. 
It is then followed by the derivation of the covariance matrix of the noise under the Gaussian window function. 
On these grounds, the excursion-set evolution of the density contrast, as the seeds of PBHs, is numerically studied in Section~\ref{sec:cnmf} after fixing and demonstrating the numerical scheme to be used, which is followed by the aforementioned findings. 
Finally, Section~\ref{sec:concl} is devoted to summarise the article.  

Natural units will be used and the convention for Fourier transform used throughout this article is fixed to be 
\begin{equation}
    \widetilde{f} (\vb*{k}) 
    = \int \dd^{3} x \, e^{- i \vb*{k} \cdot \vb*{x}} f (\vb*{x}) 
    \,\, , 
    \qquad 
    f (\vb*{x}) 
    = \int \frac{\dd^{3} k}{ (2 \pi)^{3} } \, e^{i \vb*{k} \cdot \vb*{x}} \widetilde{f} (\vb*{k}) 
    \,\, . 
    \label{eq:def_FT}
\end{equation}

\section{Standard excursion-set theory}
\label{sec:stexs}

This section reviews the conventional discussion of the excursion-set method, in which the variance of the coarse-grained density contrast is a monotonic function with respect to the coarse-graining scale. 
Due to this one-to-one correspondence, the variance can be used as the ``time'' variable. 
Section~\ref{subsec:smdc_ts} considers the Fourier-space Heaviside window function, with which the coarse-grained density contrast exhibits Markov evolution when the coarse-graining scale is variated. 
The extension to coloured noises is briefly mentioned in Section~\ref{subsec:smdc_gen}. 

\subsection{Coarse-grained density contrast}

The coarse-graining procedure defines the coarse-grained, or smoothed, density contrast $\delta (\vb*{x}, \, R)$ and relates the characteristic scale $R$ and the mass $M$ of a virialised object. 
A window function $\mathrm{W} (\abs{ \vb*{x} } / R)$ is used for this purpose, and is convolved with the density contrast $\delta (\vb*{x})$ to define 
\begin{equation}
    \delta (\vb*{x}, \, R) 
    \equiv \int \dd^{3} y \, 
    \mathrm{W} \qty( \frac{ \abs{ \vb*{x} - \vb*{y} } }{ R } ) 
    \delta ( \vb*{y} ) 
    \,\, . 
    \label{eq:def_smdc}
\end{equation}
The window function is expected to behave $\mathrm{W} (z) \simeq 1$ for $z \ll 1$ and $\mathrm{W} (z) \simeq 0$ for $z \gg 1$. 
While $\delta (\vb*{x}, \, R)$ is the field that has a value at each location in space, the position argument $\vb*{x}$ will be suppressed hereafter, therefore $\delta (R) \equiv \delta (\vb*{x}, \, R)$, since the excursion-set method is interested in the response of the quantity to the variation of the characteristic scale $R$ at each location of interest. 
The coarse-grained density contrast (\ref{eq:def_smdc}) must vanish in the limit $R \to \infty$, so it is subjected to the condition that $\delta (R = \infty) = 0$, in which structures on all the scales are smoothed out and then nothing is left. 
As $R$ decreases from infinity, $\delta (R)$ may exceed the threshold required to form a virialised object, whose mass is determined through $R = R (M)$. 
It should be noted that this relation in general depends on the choice of the window function. 

The integral in Eq.~(\ref{eq:def_smdc}), around a reference point $\vb*{x}$, accumulates and smoothes the density contrast within the length $\abs{ \vb*{x} - \vb*{y} } \lesssim R$. 
The coarse-graining procedure over a characteristic scale $R$ can be viewed in Fourier space as the superposition of the relevant modes, 
\begin{equation}
    \delta (R) 
    = \int \dd^{3} y 
    \int \frac{\dd^{3} k_{1}}{ (2 \pi)^{3} } \, 
    e^{i \vb*{k}_{1} \cdot ( \vb*{x} - \vb*{y} )} \, 
    \widetilde{\mathrm{W}} (k_{1} R) 
    \int \frac{\dd^{3} k_{2}}{ (2 \pi)^{3} } \, 
    e^{i \vb*{k}_{2} \cdot \vb*{y}} \, 
    \widetilde{\delta} (\vb*{k}_{2}) 
    = \int \frac{\dd^{3} k}{ (2 \pi)^{3} } \, 
    e^{i \vb*{k} \cdot \vb*{x}} \, 
    \widetilde{\mathrm{W}} (k R) 
    \widetilde{\delta} (\vb*{k}) 
    \,\, . 
    \label{eq:smdc_f}
\end{equation}
Under the convention (\ref{eq:def_FT}), the integral over $y$ in Eq.~(\ref{eq:smdc_f}) gives the Dirac $\delta$-function multiplied by $(2 \pi)^{3}$, 
\begin{equation}
    \int \dd^{3} y \, e^{- i \vb*{k} \cdot \vb*{y}} 
    = (2 \pi)^{3} \delta_{\rm D}^{(3)} (\vb*{k}) 
    \,\, . 
\end{equation}

In most cases, and also in this article, the variance of the coarse-grained density contrast, 
\begin{equation}
    \expval{ [ \delta (R) ]^{2} } 
    = \int \frac{\dd^{3} k_{1}}{ (2 \pi)^{3} } 
    \int \frac{\dd^{3} k_{2}}{ (2 \pi)^{3} } \, 
    e^{i ( \vb*{k}_{1} + \vb*{k}_{2} ) \cdot \vb*{x}} 
    \widetilde{\mathrm{W}} ( k_{1} R ) 
    \widetilde{\mathrm{W}} ( k_{2} R ) 
    \expval{ \widetilde{\delta} (\vb*{k}_{1}) \widetilde{\delta} (\vb*{k}_{2}) } 
    = \int \frac{\dd k}{ k } \, 
    \qty[ \widetilde{\mathrm{W}} (k R) ]^{2} 
    \mathcal{P}_{\delta} (k) 
    \,\, , 
    \label{eq:var_d}
\end{equation}
plays a central role, which, together with $\expval{ \delta (R) } = 0$, completely determines the statistical nature of the density contrast, provided that $\delta$ is Gaussian. 
The power spectrum of the density contrast itself and its nondimensionalised version are defined through 
\begin{equation}
    \expval{ \widetilde{\delta} (\vb*{k}_{1}) \widetilde{\delta} (\vb*{k}_{2})} 
    = (2 \pi)^{3} 
    \delta_{\rm D}^{(3)} ( \vb*{k}_{1} + \vb*{k}_{2} ) 
    P_{\delta} ( k_{1} ) 
    = (2 \pi)^{3} 
    \delta_{\rm D}^{(3)} ( \vb*{k}_{1} + \vb*{k}_{2} ) 
    \frac{2 \pi^{2}}{ k_{1}^{3} } 
    \mathcal{P}_{\delta} ( k_{1} ) 
    \,\, . 
\end{equation}
For the variance $S (R) \equiv \expval{ [ \delta (R) ]^{2} }$, the distribution function of $\delta (R)$ is then given by 
\begin{equation}
    f [ \delta (R) ] 
    = \frac{1}{\sqrt{2 \pi S (R)}} \, 
    \exp \qty{ - \frac{ [ \delta (R) ]^{2} }{ 2 S (R) } } 
    \,\, . 
\end{equation}

It can be seen from Eq.~(\ref{eq:smdc_f}) that, when one varies the coarse-graining scale $R$, the accumulated $\widetilde{\delta} (\vb*{k})$ and/or its weight also vary due to the window function. 
Given that $\widetilde{\delta} (\vb*{k})$ is a stochastic variable, the response of $\delta (R)$ to the variation of $R$ exhibits a stochastic process. 
In other words, for an infinitesimal decrement of $R$, the variation of the coarse-grained density contrast, 
\begin{equation}
	\dd \delta (R) 
	\equiv \delta (R) - \delta (R - \dd R) 
	= \qty[ 
		\int \frac{\dd^{3} k}{ (2 \pi)^{3} } \, 
    		e^{i \vb*{k} \cdot \vb*{x}} 
    		\pdv{ \widetilde{\mathrm{W}} (k R) }{R} 
    		\widetilde{\delta} (\vb*{k}) 
	] \, \dd R 
	\,\, , 
\end{equation}
leads to, in the limit $\dd R \to 0$, the stochastic process of $\delta (R)$, 
\begin{equation}
    \pdv{ \delta (R) }{R} 
    = \int \frac{\dd^{3} k}{ (2 \pi)^{3} } \, 
    e^{i \vb*{k} \cdot \vb*{x}} 
    \pdv{ \widetilde{\mathrm{W}} (k R) }{R} \widetilde{\delta} (\vb*{k}) 
    \equiv \xi (R) 
    \,\, . 
    \label{eq:stoc_a}
\end{equation}
Here, the stochastic ``noise'' $\xi (R)$ has been introduced. 
The correlation properties of this noise are given by $\expval{\xi (R)} = 0$
by construction, and the variance, 
\begin{equation}
	\expval{ \xi (R_{1}) \xi (R_{2}) } 
	= \int \frac{ \dd k }{ k } 
    	\pdv{ \widetilde{\mathrm{W}} (k R_{1}) }{R_{1}} 
	\pdv{ \widetilde{\mathrm{W}} (k R_{2}) }{R_{2}} 
	\mathcal{P}_{\delta} (k).
\end{equation}
It should be noted, however, that the stochastic process (\ref{eq:stoc_a}) must be conditioned to the ``initial'' condition $\delta (R = \infty) = 0$, which is sometimes inconvenient to treat. 
The variance $S$ is rather widely used in the literature, assuming that there is a one-to-one correspondence between $R$ and $S$. 

Using the variance $S$ as the time variable, the stochastic evolution of $\delta$ is described by 
\begin{equation}
    \pdv{\delta (S)}{S} 
    = \qty[ \dv{S (R)}{R} ]^{-1} \pdv{\delta (R)}{R} 
    \equiv \xi (S) 
    \,\, , 
    \label{eq:stoc_b}
\end{equation}
where $\delta (S) \equiv \delta [ R = R (S) ]$ and $\expval{ \xi (S) } = 0$. 
The variance is assumed to be a monotonically increasing function with respect to the decrement of $R$, and $S = 0$ corresponds to $R = \infty$ in particular. 
In terms of $S$, the covariance of the noise $\xi (S)$ for a general window function is then given by 
\begin{equation}
    \expval{ \xi (S_{1}) \xi (S_{2}) } 
    = \frac{
        \displaystyle 
        \int_{0}^{\infty} \frac{\dd k}{k} \, 
        \pdv{ \widetilde{\mathrm{W}} (k R_{1}) }{ R_{1} } 
        \pdv{ \widetilde{\mathrm{W}} (k R_{2}) }{ R_{2} } 
        \mathcal{P}_{\delta} (k) 
    }{
        \displaystyle 
        \qty{ 
            \int_{0}^{\infty} \frac{\dd k}{k} \, 
            \pdv{R_{1}} \qty[ \widetilde{\mathrm{W}} (k R_{1}) ]^{2}  
            \mathcal{P}_{\delta} (k) 
        } 
        \qty{ 
            \int_{0}^{\infty} \frac{\dd k}{k} \, 
            \pdv{R_{2}} \qty[ \widetilde{\mathrm{W}} (k R_{2}) ]^{2} 
            \mathcal{P}_{\delta} (k) 
        } 
    } 
    \,\, . 
    \label{eq:covar_xis}
\end{equation}
Each factor on the right-hand side of Eq.~(\ref{eq:covar_xis}) should be understood as a function of $S_{1} = S_{1} (R_{1})$ and $S_{2} = S_{2} (R_{2})$. 

The excursion-set method solves Eq.~(\ref{eq:stoc_b}), in which $\delta (S)$ exhibits a stochastic process subjected to the stochastic noise $\xi (S)$ and to the initial condition $\delta (S = 0) = 0$. 
It is understood that if $\delta (S)$ exceeds a certain given threshold $\delta_{\rm th}$ at a scale $R$, a virialised object is formed at $\vb*{x}$ with its mass corresponding to $R$. 

One of the advantages of the excursion-set method is that it provides a natural solution to the so-called cloud-in-cloud problem, which is present in the Press--Schechter formalism. 
In the Press--Schechter formulation, configurations in which a smaller virialised region is absorbed by a larger one are not accounted for. 
In such cases, even if both $\delta (S_{1}) > \delta_{\rm th}$ \textit{and} $\delta (S_{2}) > \delta_{\rm th}$ hold, the outcome would be a single virialised object associated with the larger scale. 
The excursion-set method resolves this issue by identifying the \textit{largest} coarse-graining scale at which the coarse-grained density contrast first exceeds the threshold. 
Since it corresponds to the \textit{smallest} variance, the analysis reduces to deriving the first-passage-time distribution of the stochastic process (\ref{eq:stoc_b}). 

For consistency in later sections, let us introduce $\tau \equiv 1/R$ and rewrite the above covariance in terms of $\tau$. 
To do so, the stochastic noise in terms of $\tau$ is introduced in a similar way as Eq.~(\ref{eq:stoc_b}), by 
\begin{equation}
    \pdv{\delta (\tau)}{\tau} 
    = \eval{ 
        \int \frac{\dd^{3} k}{(2 \pi)^{3}} 
        \, e^{i \vb*{k} \cdot \vb*{x}} 
        \qty( - \frac{k}{\tau^{2}} ) 
        \pdv{ \widetilde{\mathrm{W}} (z) }{z} 
        \widetilde{\delta} (\vb*{k}) 
    }_{z = k / \tau} 
    \equiv \xi (\tau) 
    \,\, . 
\end{equation}
The covariance matrix in terms of $\tau$ is then given by 
\begin{equation}
    \expval{ \xi (\tau_{1}) \xi (\tau_{2}) } 
    = \eval{ 
        \frac{1}{\tau_{1}^{2} \tau_{2}^{2}} 
        \int \frac{\dd k}{k} \, k^{2} 
        \pdv{ \widetilde{\mathrm{W}} (z_{1})}{z_{1}} 
        \pdv{ \widetilde{\mathrm{W}} (z_{2})}{z_{2}} 
        \mathcal{P}_{\delta} (k) 
    }_{z_{1} = k / \tau_{1}, \, z_{2} = k / \tau_{2}} 
    \,\, . \label{eq:ncov_smdc}
\end{equation}

\subsection{Heaviside window function as touchstone}
\label{subsec:smdc_ts}

A widely studied window function in the context of the excursion-set method is the Heaviside function in Fourier space, 
\begin{equation}
	\widetilde{\rm W} (z) = \Theta (1 - z) 
	\,\, , 
	\label{eq:smdc_fH}
\end{equation}
For Eq.~(\ref{eq:smdc_fH}), $\widetilde{\rm W} (k R)$ extracts the long-wavelength modes that meet $k R \leq 1$, while the short-wavelength modes with $k R \geq 1$ are discarded. 

In this case, it can be confirmed that the variation of the coarse-grained density contrast with respect to the coarse-graining scale $R$ exhibits the Brownian motion driven by the uncorrelated noise, 
\begin{equation}
	\pdv{ \delta (S) }{ S } 
	= \xi (S) 
	\,\, , 
	\qquad 
	\begin{cases}\begin{aligned}
		\expval{ \xi (S) } &= 0 \,\, , 
		\\ 
		\expval{ \xi (S_{1}) \xi (S_{2}) } &= \delta_{\rm D} (S_{1} - S_{2}) 
		\,\, . 
	\end{aligned}\end{cases}
	\label{eq:smdc_wbw}
\end{equation}
The variance of $\delta (S)$ can be derived directly from Eq.~(\ref{eq:smdc_wbw}), or also directly from Eq.~(\ref{eq:smdc_f}), 
\begin{equation}
	\expval{ \delta (S_{1}) \delta (S_{2}) } 
	= \int_{0}^{S_{1}} \dd s_{1} \int_{0}^{S_{2}} \dd s_{2} \, \expval{ \xi (s_{1}) \xi (s_{2}) } 
	= \int_{0}^{1 / \mathrm{max} (R_{1}, \, R_{2})} \frac{ \dd k }{k} \, \mathcal{P}_{\delta} (k) 
	\,\, . 
\end{equation}
In the ``equal-time'' limit, it reduces to the variance $S = \expval{ [ \delta (S) ]^{2} }$ defined in Eq.~(\ref{eq:var_d}). 

The fact that $\delta (S)$ evolves stochastically governed by Eq.~(\ref{eq:smdc_wbw}) motivates one to consider the probability distribution function of $\delta (S)$, which follows the diffusion equation, 
\begin{equation}
	\pdv{f}{S} = \frac{1}{2} \pdv[2]{f}{\delta} 
	\,\, , 
	\qquad 
	f = f (\delta, \, S) = f (\delta, \, S \mid \delta_{0}, \, S_{0}) 
	\,\, . 
	\label{eq:smdc_dfeq}
\end{equation}
The diffusion equation (\ref{eq:smdc_dfeq}) must be supplemented by the initial and two boundary conditions. 
Under the initial condition $f (\delta, \, S) = \delta_{\rm D} (\delta)$ at $S = 0$, the free solution to Eq.~(\ref{eq:smdc_dfeq}) is given by the Gaussian distribution, $f (\delta, \, S) = (1 / \sqrt{2 \pi S}) \exp \, ( - \delta^{2} / 2 S )$. 
In practice, an absorbing boundary is placed at $\delta = \delta_{\rm th}$. 
The solution to that situation can be written down by virtue of the method of images, 
\begin{equation}
	f (\delta, \, S) 
	= \frac{1}{ \sqrt{2 \pi S} } \qty{ 
		\exp \qty( - \frac{ \delta^{2} }{ 2 S } ) 
		- \exp \qty[ - \frac{ ( \delta - 2 \delta_{\rm th} )^{2} }{ 2 S } ] 
	} 
	\,\, . 
	\label{eq:smdc_sdsol}
\end{equation}
The first-passage-time distribution of $S$, denoted $f_{\rm FPT} (\delta, \, S)$, follows from Eq.~(\ref{eq:smdc_sdsol}) through the survival probability. 
The quantity $f_{\rm FPT} (\delta, \, S)$ tells us the distribution of $S$ at which the coarse-grained density contrast pierces the threshold, given the stochastic realisations started from $\delta$ and evolved under the noise. 
Therefore, the integration of $f_{\rm FPT} (\delta, \, S)$ over $S$ gives the fraction of all the realisations that, until the time $S$, they have already pierced the threshold. 
To put it in the converse way, the complementary fraction corresponds to the still-surviving realisations. 
One then notices that 
\begin{equation}
	\int_{- \infty}^{\delta_{\rm th}} \dd \delta \, \pdv{ f (\delta, \, S) }{ \delta } 
	= 1 - \int_{0}^{S} \dd s \, f_{\rm FPT} (\delta, \, s) 
	\,\, . 
	\label{eq:smdc_sv}
\end{equation}
Differentiation of Eq.~(\ref{eq:smdc_sv}) with respect to $S$ and then using the diffusion equation~(\ref{eq:smdc_dfeq}) together with its solution~(\ref{eq:smdc_sdsol}), one arrives at the first-passage-time distribution, 
\begin{equation}
	f_{\rm FPT} (S) 
	= - \frac{1}{2} \eval{ 
		\pdv{ f (\delta, \, S) }{ \delta } 
	}_{\delta = -\infty }^{\delta = \delta_{\rm th}} 
	=  \frac{ \delta_{\rm th} }{ \sqrt{2 \pi} \, S^{3/2} } \exp \biggl( - \frac{ \delta_{\rm th}^{2} }{ 2 S } \biggr) 
	\,\, . 
	\label{eq:smdc_fc}
\end{equation}
It should be noted that, in Eq.~(\ref{eq:smdc_fc}), there are the contributions from the first (source) and second (image) terms in Eq.~(\ref{eq:smdc_sdsol}). 
Due to the use of the Fourier-space Heaviside window function, or equivalently due to the Markov property of the noise, both the source and image terms contribute equally, giving rise to the factor two in Eq.~(\ref{eq:smdc_fc}), which is though compensated by $1/2$. 
This is why the so-called fudge factor two naturally arises and is resolved in the excursion-set method, for the particular choice of the window function given by Eq.~(\ref{eq:smdc_fH}). 

Though there is no analytical result in general, Eq.~(\ref{eq:smdc_sdsol}) can be generalised to situations where correlated noises play a role. 
In such cases, one defines the formation probability, schematically written as (see \textit{e.g.}~Ref.~\cite{Kushwaha:2025zpz}) 
\begin{equation}
	P (S) 
	= P_{1} [ \mathrm{A} (S) ] + P_{2} [ \mathrm{B} (S) ] 
	\,\, , 
    \label{eq:smdc_fgenp}
\end{equation}
where $A (S) \equiv \qty{ \delta \mid \delta (S) > \delta_{\rm th} }$ and $B (S) \equiv \qty{ \delta \mid \delta (S) < \delta_{\rm th} \cap {}^{\exists} S' < S ~\text{s.t.}~ \delta (S') > \delta_{\rm th} }$. 
In the presence of correlated noises, the relation that $P_{1} = P_{2}$ no longer holds, \textit{i.e.}~source and image contribute in a different way. 

\paragraph*{\textit{Power-law power spectrum.}}
To observe the difference in the evolution of the variance of $\delta$, between the cases with the uncorrelated and correlated noises, let us consider the power-law power spectrum, $\mathcal{P}_{\delta} (k) = \mathcal{P}_{0} ( k / k_{\star} )^{\alpha}$ with $\alpha > 0$ and a fiducial scale $k_{\star}$. 
The variance (\ref{eq:var_d}) is written as 
\begin{align}
	S (R) 
	&= \int_{0}^{\infty} \frac{\dd z}{z} \, \Theta (1 - z) \mathcal{P}_{0}
    \qty( \frac{k}{k_{\star}} )^{\alpha}
	= \frac{\mathcal{P}_{0}}{\alpha}  \frac{1}{\qty(k_{\star} R)^{\alpha}}
	\,\, , 
	\label{eq:st_var_eps}
\end{align}
and the stochastic process is given by  (\ref{eq:smdc_wbw}). 
The relation (\ref{eq:st_var_eps}) can be inverted to give $k_{\star} R = [\mathcal{P}_{0} / \alpha S (R) ]^{1/\alpha}$. 

\subsection{Smooth window function and correlated noise}
\label{subsec:smdc_gen}

When a window function other than Eq.~(\ref{eq:smdc_fH}) is used, the noise covariance becomes correlated, or \textit{coloured}, in general. 
In such cases, the Markov property of the stochastic process no longer holds, so there exists neither the simplified diffusion equation for the distribution of $\delta$, such as Eq.~(\ref{eq:smdc_dfeq}), nor the analytical expression for the first-passage-time distribution, such as Eq.~(\ref{eq:smdc_fc}). 
One is then led to conduct numerical simulations. 

There is literature that considers the correlated noise in the standard excursion-set method, see \textit{e.g.}~Refs.~\cite{Bond:1990iw, Lacey:1994su, Musso:2013pha, Albrecht:1982wi}. 
Though simulations of the excursion-set method in a non-PBH context are not repeated here, let us briefly mention to a case where the increment of the variance is altered by a smooth window function. 

A convenient choice of the window function that demonstrates the correlated scales is the Gaussian window function, 
\begin{equation}
    \widetilde{\mathrm{W}} (z) 
    = \exp \qty( - \frac{z^{2}}{2} ) 
    \,\, . 
    \label{eq:gf_wf}
\end{equation}
The convention of Eq.~(\ref{eq:gf_wf}) follows the one in Ref.~\cite{Ando:2018qdb}. 
Under the power-law power spectrum of $\delta$, the variance reads 
\begin{equation}
    S(R) 
    = \frac{\mathcal{P}_{0}}{(k_{\star} R)^{\alpha}} 
    \int_{0}^{\infty} \frac{\dd z}{z} \, z^{\alpha} \exp\, ( - z^{2} ) 
    = \frac{\mathcal{P}_{0}}{2}\frac{1}{(k_{\star} R)^{\alpha}}  \Gamma \qty( \frac{\alpha}{2}) 
	\,\, .
	\label{eq:st_var_eps_Gauss}
\end{equation} 
One finds that by the use of a smooth window function the variance remains unchanged except the prefactor $\Gamma (\alpha / 2)$. 
However, this is not the case for the covariance matrix of the noise that drives the stochastic behaviour of the coarse-grained density contrast. 
Indeed, it reads 
\begin{equation}
    \expval{ \xi (S_{1}) \xi (S_{2}) } 
    = \frac{2^{\alpha / 2 +1}}{ \mathcal{P}_{0} } 
    \frac{ \alpha / 2 + 1 }{ \Gamma (\alpha / 2 + 1) } 
    \frac{ (k_{\star} R_{1})^{\alpha + 2} (k_{\star} R_{2})^{\alpha + 2} }{ [ (k_{\star} R_{1})^{2} + (k_{\star} R_{2})^{2} ]^{\alpha / 2 + 2} } 
    = 2^{\alpha / 2} \frac{\alpha + 2}{\alpha} 
    \frac{ (S_{1} S_{2})^{2 / \alpha} }{ ( S_{1}^{2 / \alpha} + S_{2}^{2 / \alpha} )^{\alpha / 2 + 2} } 
    \,\, . 
    \label{eq:gf_lan_cov}
\end{equation}
The absence of the Dirac-$\delta$ function in the covariance matrix can be seen, which is a crucial difference from the simplest case explained in Section~\ref{subsec:smdc_ts}. 
From Eq.~(\ref{eq:gf_lan_cov}), it can be seen that the variance of the noise reads $\expval{[\xi(S)]^2} = [ (\alpha + 2) / 4 \alpha ] / S$, so that it decreases in the large-$S$ limit in proportion to $1 / S$, contrary to Eq.~(\ref{eq:smdc_wbw}). 

\section{Excursion-set theory for PBH formation}
\label{sec:pbexs}

An underlying assumption made in the standard excursion-set method reviewed in Section~\ref{sec:stexs} is that the variance of $\delta (R)$ is a monotonic function with respect to the coarse-graining scale. 
This ceases to hold in the context of the formation of PBHs, due to the crucial difference that a PBH may be formed when the scale of interest reenters the horizon, whereas non-PBH objects such as halos can be formed at an arbitrary sub-horizon scale. 

The corresponding stochastic process is then driven by the correlated noise, as was pointed out in Ref.~\cite{Kushwaha:2025zpz} and will be reviewed here, deriving all the relevant formulas in a detailed manner. 
Based on the general equations given in Section~\ref{subsec:pbexs_gen}, the covariance matrix of the noise will analytically be derived in Section~\ref{subsec:pbexs_ncov} for the two kinds of window functions, by which one is ready to numerically simulate the coloured-noise dynamics in Section~\ref{sec:cnmf}. 

\subsection{General formulas}
\label{subsec:pbexs_gen}

On super-horizon scales, the curvature perturbation $\zeta$ is conserved and is related to the density contrast. 
At the lowest order in the gradient expansion~\cite{Salopek:1990jq, Deruelle:1994iz, Shibata:1999zs, Lyth:2004gb}, the relation between those quantities is given by~\cite{Harada:2015yda}
\begin{equation}
    \delta (\vb*{x}, \, t) 
    = - \frac{2 (w + 1)}{3 w + 5} \frac{1}{ ( a H )^{2} } 
    \exp \qty[ - \frac{5}{2} \zeta (\vb*{x}) ] 
    \nabla^{2} \exp \qty[ \frac{ \zeta (\vb*{x}) }{2} ] 
    \simeq - \frac{4}{9} \frac{1}{(a H)^{2}} \nabla^{2} \zeta (\vb*{x})  
    \,\, ,
    \label{eq:g_dlin}
\end{equation}
on comoving slices. 
The radiation-dominated universe is assumed hereafter, that is, the equation-of-state parameter is fixed to be $w = 1/3$. 
The relation is linearised in Eq.~(\ref{eq:g_dlin}) by keeping the leading-order contribution in $\zeta$, following the conventional studies, though the non-linear effect would become relevant when one wants to conduct more careful analyses~\cite{Musco:2018rwt, Escriva:2019phb, Young:2019yug, DeLuca:2019qsy, Germani:2019zez, Kawasaki:2019mbl, Germani:2023ojx}. 
The relation (\ref{eq:g_dlin}) states that, even though the curvature perturbation is conserved on super-horizon scales, the density contrast is a function of time, where $1 / aH$ is the comoving horizon scale. 

What characterises a PBH is that the mass scale is determined solely by the horizon scale.\footnote{
	The order-unity factor $\gamma$ defined as the ratio of a PBH to the horizon mass is not considered throughout this article. 
}  
The coarse-graining procedure is performed in the same way as Eq.~(\ref{eq:def_smdc}), allocating the mass for a PBH associated with the enhanced density contrast. 
However, due to the identification of the horizon scale with the coarse-graining scale (or equivalently the mass scale), $R = 1 / aH$ in Eq.~(\ref{eq:g_dlin}), the counterpart of Eq.~(\ref{eq:def_smdc}) in the context of the formation of PBHs reads 
\begin{equation}
    \delta(\vb*{x}, \, R) 
    \equiv \int \dd^{3} y \, \mathrm{W} \qty( \frac{ \abs{ \vb*{x} - \vb*{y} } }{R} ) \delta (\vb*{y}, \, t) 
    = - \frac{4}{9} R^{2} 
    \int \dd^{3} y \, \mathrm{W} \qty( \frac{ \abs{ \vb*{x} - \vb*{y} } }{R} ) \nabla^{2} \zeta (\vb*{y}) 
    \,\, . 
    \label{eq:g_smdens}
\end{equation}
The notation in the left-hand side in Eq.~(\ref{eq:g_smdens}) is due to the fact that the coarse-graining scale $R$, the horizon scale $1 / aH$, and the time $t$ can be used interchangeably. 
One sees that the characteristic nature of PBHs introduces the factor $R^{2}$ in front of the integral, which results in the additional contribution to the covariance of the noise. 

In most inflationary scenarios that realise the formation of PBHs, the power spectrum is amplified on small scales, while on large scales it remains scale-invariant. 
In other words, the relevant part of the power spectrum for the formation of PBHs is typically localised, meaning that there no longer exists a one-to-one correspondence between the coarse-graining scale and the variance of the coarse-grained density contrast. 
This is why the ``time'' variable $\tau \equiv 1 / R$ is introduced instead of the variance. 
In addition to this, the position argument $\vb*{x}$ will again be suppressed, with a reference location being focused hereafter. 
Writing $\delta ( \vb*{x}, \, R ) = \delta (\tau)$, Eq.~(\ref{eq:g_smdens}) becomes 
\begin{equation}
    \delta (\tau) 
    = - \frac{4}{9} \frac{1}{\tau^{2}} 
    \int \dd^{3} y \, \mathrm{W} ( \abs{ \vb*{x} - \vb*{y} } \, \tau ) \nabla^{2} \zeta (\vb*{y}) 
    = \frac{4}{9} 
    \int \frac{\dd^{3} k}{ (2 \pi)^{3} } 
    \, e^{i \vb*{k} \cdot \vb*{x}} \, 
    \widetilde{\zeta} (\vb*{k}) 
        \qty[ 
            z^{2} \widetilde{\mathrm{W}} (z) 
        ]_{z = k / \tau} 
    \,\, . 
\end{equation}

To derive the stochastic process exhibited by $\delta (\tau)$, let us differentiate $\delta (\tau)$ with respect to $\tau$, to obtain  
\begin{equation}
    \pdv{\delta (\tau)}{\tau} 
    = \frac{8}{9} \frac{1}{\tau^{3}} \int \dd^{3} y \, \mathrm{W} ( \abs{ \vb*{x} - \vb*{y} } \, \tau ) \nabla^{2} \zeta (\vb*{y}) 
    - \frac{4}{9} \frac{1}{\tau^{2}} \int \dd^{3} y \, \pdv{ \mathrm{W} (\abs{ \vb*{x} - \vb*{y} } \, \tau) }{ \tau } \nabla^{2} \zeta (\vb*{y}) 
    \,\, . 
    \label{eq:g_derd}
\end{equation}
The second term in Eq.~(\ref{eq:g_derd}), in which the derivative hits the window function, also appears in the standard excursion-set formulation reviewed in the previous section. 
There, it is the only source that drives the stochastic variation of the coarse-grained density contrast. 
What is peculiar to the context surrounding PBH is the first term, originating from the identification of the mass scale with the horizon scale. 
As will be seen below, this term gives rise to the correlated noise, even when the Fourier-space Heaviside window function is employed. 

To proceed, it is more convenient to rewrite Eq.~(\ref{eq:g_derd}) in Fourier integral, 
\begin{equation}
    \pdv{ \delta (\tau) }{\tau} 
    = \xi (\tau) 
    \equiv \frac{4}{9} 
    \int \frac{\dd^{3} k}{ (2 \pi)^{3} } 
    \, e^{i \vb*{k} \cdot \vb*{x}} \, 
    \widetilde{\zeta} (\vb*{k}) 
    \qty( - \frac{ k }{\tau^{2}} ) 
    \qty[ 
        2 z \widetilde{\mathrm{W}} (z) + z^{2} \pdv{ \widetilde{\mathrm{W}} (z) }{z} 
    ]_{z = k / \tau} 
    \,\, . 
    \label{eq:g_lano}
\end{equation}
The stochastic noise $\xi (\tau)$ has been introduced here. 
Since the zero-mean of the curvature perturbation is assumed, the stochastic mean of $\xi$ also vanishes as well, $\expval{ \xi (\tau) } = 0$, whereas the covariance matrix of the noise reads 
\begin{align}
    \expval{ \xi (\tau_{1}) \xi (\tau_{2}) } 
    &= \frac{16}{81} 
    \int \frac{\dd k}{ k } 
    \mathcal{P}_{\zeta} ( k ) 
    \frac{ k^{2} }{ \tau_{1}^{2} \tau_{2}^{2} } 
    \times \left[ 
        4 z_{1} z_{2} \widetilde{\mathrm{W}} (z_{1}) \widetilde{\mathrm{W}} (z_{2}) 
        + 2 z_{1} z_{2}^{2} \widetilde{\mathrm{W}} (z_{1}) \pdv{ \widetilde{\mathrm{W}} (z_{2}) }{ z_{2} } 
    \right. 
    \notag \\ 
    &\quad \left. 
        + \, 2 z_{2} z_{1}^{2} \widetilde{\mathrm{W}} (z_{2}) \pdv{ \widetilde{\mathrm{W}} (z_{1}) }{ z_{1} } 
        + z_{1}^{2} z_{2}^{2} \pdv{ \widetilde{\mathrm{W}} (z_{1}) }{ z_{1} } \pdv{ \widetilde{\mathrm{W}} (z_{2}) }{ z_{2} } 
    \right]_{z_{1} = k / \tau_{1}, \, z_{2} = k / \tau_{2}} .
	\label{eq:ncov_gen}
\end{align}
It can be noticed that, comparing Eq.~(\ref{eq:ncov_gen}) with Eq.~(\ref{eq:ncov_smdc}), all the terms but the last one in the square bracket are peculiar to the formation of PBHs. 
When a smooth window function is implemented, all four terms give rise to the correlated noise. 
When the Fourier-space Heaviside window function is used, on the other hand, all the terms that include $\widetilde{\mathrm{W}}$, not the derivative of $\widetilde{\mathrm{W}}$, introduce the correlation over the scales. 
A demonstration of the latter case was presented in Ref.~\cite{Kushwaha:2025zpz}, to which another one with a smooth window function is supplemented in Section~\ref{subsec:ncov_g}. 
The variance of the noise is derived in the $\tau_{1} = \tau_{2} \equiv \tau$ limit, 
\begin{subequations}
\begin{align}
    \expval{ \qty[ \xi (\tau) ]^{2} } 
    &= \frac{16}{81} 
    \int \frac{\dd k}{ k } 
    \mathcal{P}_{\zeta} ( k ) 
    \frac{ k^{2} }{ \tau^{4} } 
    \qty{ 
        4 z^{2} \qty[ \widetilde{\mathrm{W}} (z) ]^{2} 
        + 4 z^{3} \widetilde{\mathrm{W}} (z) \pdv{ \widetilde{\mathrm{W}} (z) }{z} 
        + z^{4} \qty[ \pdv{ \widetilde{\mathrm{W}} (z) }{z} ]^{2} 
    }_{z = k / \tau} 
    \\ 
    &= \frac{16}{81} 
    \int_{0}^{\infty} \frac{\dd z}{ z } 
    \mathcal{P}_{\zeta} ( \tau z ) 
    \frac{ z^{2} }{ \tau^{2} } 
    \qty{ 
        4 z^{2} \qty[ \widetilde{\mathrm{W}} (z) ]^{2} 
        + 4 z^{3} \widetilde{\mathrm{W}} (z) \pdv{ \widetilde{\mathrm{W}} (z) }{z} 
        + z^{4} \qty[ \pdv{ \widetilde{\mathrm{W}} (z) }{z} ]^{2} 
    } 
    \,\, . 
\end{align}
\end{subequations}
From the stochastic process, $\dd \delta (\tau) / \dd \tau = \xi (\tau)$, the covariance matrix and the variance of $\delta$ can also be obtained, 
\begin{subequations}
	\begin{align}
	\expval{ \delta (\tau_{1}) \delta (\tau_{2}) } 
	&= \int_{0}^{\tau_{1}} \dd \widehat{\tau}_{1} 
	\int_{0}^{\tau_{2}} \dd \widehat{\tau}_{2} 
	\expval{ \xi (\widehat{\tau}_{1}) \xi (\widehat{\tau}_{2}) } 
	= \frac{16}{81} \int \frac{\dd k}{k} \, \mathcal{P}_{\zeta} (k) z_{1}^{2} z_{2}^{2} \widetilde{\mathrm{W}} (z_{1}) \widetilde{\mathrm{W}} (z_{2}) 
	\,\, , 
	\\ 
	\expval{ \qty[ \delta (\tau) ]^{2} } 
    	&= \frac{16}{81} 
    	\int_{0}^{\infty} \frac{\dd z}{ z } 
    	\mathcal{P}_{\zeta} ( \tau z )   \qty[ z^{2} \widetilde{\mathrm{W}} (z) ]^2 
	\,\, .
	\end{align}
\end{subequations}
It is noted that $\delta (\tau = 0) = 0$, provided that $\tau = 0$ corresponds to $R = \infty$. 
It should also be mentioned that the variance of $\delta$ should not behave badly as $\tau \to 0$, \textit{i.e.}~$\expval{ [ \delta (\tau = 0) ]^{2} } = 0$ should hold due to the aforementioned ``initial'' condition. 

\subsection{Noise covariance}
\label{subsec:pbexs_ncov}

It is necessary to fix the functional forms of both the power spectrum of the curvature perturbation, $\mathcal{P}_{\zeta} (k)$, and the window function $\widetilde{\mathrm{W}} (z)$, in order to proceed further. 
For the power spectrum, there are various functions that have been assumed in the literature, from the Dirac $\delta$-function (as the narrow limit) to the scale-invariant (broad) function. 
The Dirac $\delta$-function enables one to derive quantities of interest such as the mass function analytically but it may be too simplified.
The broad power spectrum is however sometimes inconvenient in practice, since the cutoff scales may be required to ensure the convergence of the statistical quantities. 
With regard to window functions, either the Fourier-space Heaviside window function, the Gaussian, or the real-space Heaviside window function is typically used in the literature. 

Hereafter, the power spectrum is assumed to be localised\footnote{
	See for instance Refs.~\cite{Germani:2018jgr, MoradinezhadDizgah:2019wjf, DeLuca:2020ioi, Yoo:2020dkz, Sureda:2020vgi, Banerjee:2024nkv} for studies on broad power spectra. 
}
around a fiducial scale $k_{\star}$. 
In particular, it is fixed to be the Gaussian power spectrum with the broken power-law factor, given by 
\begin{equation}
	\mathcal{P}_{\zeta} (k) 
	= \frac{ \mathcal{P}_{0} k_{\star} }{ \sqrt{2 \pi} \, \Delta } 
    \qty( \frac{k}{k_{\star}} )^{\alpha} 
    \exp \qty[ - \frac{ (k - k_{\star})^{2} }{ 2 \Delta^{2} } ] 
    \,\, . 
    \label{eq:psG}
\end{equation}
The choice of Eq.~(\ref{eq:psG}) enables us to derive the covariance matrix of the noise analytically, for the two window functions, $\widetilde{\mathrm{W}} (z) = \Theta (1 - z)$ and $\widetilde{\mathrm{W}} (z) = \exp \, ( - z^{2} / 2 )$. 
The power-law part, $( k / k_{\star} )^{\alpha}$, is required to satisfy the initial condition imposed at $\tau = 0$, as will be seen. 
The overall amplitude of the power spectrum is controlled by $\mathcal{P}_{0}$. 
It is indeed localised around $k = k_{\star}$, with the standard deviation being approximately specified by $\Delta$, while $\alpha > 0$ shifts $\mathrm{argmax} \, \mathcal{P}_{\zeta} (k)$ from $k_{\star}$ towards the small-scale (\textit{i.e.}~larger $k$) direction, at $( k_{\star} + \sqrt{ k_{\star}^{2} + 4 \alpha \Delta^{2} } ) / 2$. 

With the power spectrum (\ref{eq:psG}), the noise covariance matrix (\ref{eq:ncov_gen}) reduces to 
\begin{align}
	\Delta^{2} \expval{ \xi (\tau_{1}) \xi (\tau_{2}) } 
	&= \frac{16}{81} 
	\frac{ \mathcal{P}_{0} }{ \sqrt{2 \pi} } \frac{1}{\widetilde{k}_{\star}^{\alpha - 1}} 
	\frac{1}{ \widetilde{\tau}_{1}^{2} \widetilde{\tau}_{2}^{2} } 
	\int_{0}^{\infty} \frac{\dd x}{x} \, x^{\alpha + 2} 
	\exp \qty[ - \frac{ ( x - \widetilde{k}_{\star} )^{2} }{2} ] 
	\notag \\ 
	&\quad \times \left\{ 
		4 z_{1} z_{2} \widetilde{\mathrm{W}} (z_{1}) \widetilde{\mathrm{W}} (z_{2})  
        	+ \, 2 z_{1} z_{2}^{2} \widetilde{\mathrm{W}} (z_{1}) \dv{ \widetilde{\mathrm{W}} (z_{2}) }{ z_{2} } 
	\right. 
	\notag \\ 
	&\quad \quad~~ \left. 
		+ \, 2 z_{1}^{2} z_{2} \dv{ \widetilde{\mathrm{W}} (z_{1}) }{ z_{1} } \widetilde{\mathrm{W}} (z_{2}) 
        		+ z_{1}^{2} z_{2}^{2} \dv{ \widetilde{\mathrm{W}} (z_{1}) }{ z_{1} } \dv{ \widetilde{\mathrm{W}} (z_{2}) }{ z_{2} }
	\right\}_{z_{1} = x / \widetilde{\tau}_{1}, \, z_{2} = x / \widetilde{\tau}_{2}} 
	\,\, .
	\label{eq:ncovG_gen_a}
\end{align}
It should be noted that the product $\Delta \cdot \xi (\tau)$ has no dimension. 
For our computational convenience and for the numerical implementation in the next section, the nondimensionalised quantities have been introduced, 
\begin{equation}
	\widetilde{\tau}_{i} 
    \equiv \frac{ \tau_{i} }{ \Delta } 
	\,\, , 
	\qquad 
	\widetilde{k}_{\star} 
    \equiv \frac{ k_{\star} }{ \Delta } 
	\,\, . 
	\label{eq:ndimv}
\end{equation}
As was mentioned previously, there are essentially three contributions in Eq.~(\ref{eq:ncovG_gen_a}), which motivates us to define 
\begin{subequations}
	\label{eq:ncovG_igen}
	\begin{align}
		I_{1} (\widetilde{\tau}_{1}, \, \widetilde{\tau}_{2}) 
		&\equiv \frac{4}{\widetilde{\tau}_{1}^{3} \widetilde{\tau}_{2}^{3}} 
		\times \int_{0}^{\infty} \frac{\dd x}{x} \, x^{\alpha + 4} 
		\exp \qty[ - \frac{ (x - \widetilde{k}_{\star})^{2} }{2} ] 
		\widetilde{\mathrm{W}} (z_{1}) \widetilde{\mathrm{W}} (z_{2}) 
		\eval{}_{z_{1} = x / \widetilde{\tau}_{1}, \, z_{2} = x / \widetilde{\tau}_{2}} 
		\,\, , 
		\label{eq:ncovG_i1gen}
		\\ 
		I_{2} (\widetilde{\tau}_{1}, \, \widetilde{\tau}_{2}) 
		&\equiv \frac{-2}{\widetilde{\tau}_{1}^{3} \widetilde{\tau}_{2}^{4}} 
		\times \int_{0}^{\infty} \frac{\dd x}{x} \, x^{\alpha + 5} 
		\exp \qty[ - \frac{ (x - \widetilde{k}_{\star})^{2} }{2} ] 
		\widetilde{\mathrm{W}} (z_{1}) \dv{\widetilde{\mathrm{W}} (z_{2})}{z_{2}} 
		\eval{}_{z_{1} = x / \widetilde{\tau}_{1}, \, z_{2} = x / \widetilde{\tau}_{2}}  
		\,\, , 
		\label{eq:ncovG_i2gen}
		\\ 
		I_{3} (\widetilde{\tau}_{1}, \, \widetilde{\tau}_{2}) 
		&\equiv \frac{1}{\widetilde{\tau}_{1}^{4} \widetilde{\tau}_{2}^{4}} 
		\times \int_{0}^{\infty} \frac{\dd x}{x} \, x^{\alpha + 6} 
		\exp \qty[ - \frac{ (x - \widetilde{k}_{\star})^{2} }{2} ] 
		\dv{\widetilde{\mathrm{W}} (z_{1})}{z_{1}}  \dv{\widetilde{\mathrm{W}} (z_{2})}{z_{2}} 
		\eval{}_{z_{1} = x / \widetilde{\tau}_{1}, \, z_{2} = x / \widetilde{\tau}_{2}} 
		\,\, , 
		\label{eq:ncovG_i3gen}
	\end{align}
\end{subequations}
so that Eq.~(\ref{eq:ncovG_gen_a}) can be expressed, in terms of these functions, as follows, 
\begin{equation}
	\Delta^{2} \expval{ \xi (\tau_{1}) \xi (\tau_{2}) } 
	= \frac{16}{81} \frac{\mathcal{P}_{0}}{ \sqrt{2 \pi} } 
	\frac{1}{\widetilde{k}_{\star}^{\alpha - 1}} 
	\qty{
		I_{1} (\widetilde{\tau}_{1}, \, \widetilde{\tau}_{2}) 
		- \qty[ 
			I_{2} (\widetilde{\tau}_{1}, \, \widetilde{\tau}_{2}) + I_{2} (\widetilde{\tau}_{2}, \, \widetilde{\tau}_{1}) 
		] 
		+ I_{3} (\widetilde{\tau}_{1}, \, \widetilde{\tau}_{2}) 
	} 
	\,\, . 
	\label{eq:ncovG_gen_b}
\end{equation}
The remaining things, for the concrete expression of the noise covariance, are to specify the window function and then to perform the integrals defined in Eqs.~(\ref{eq:ncovG_igen}). 
To bring out the distinctive features in the mass function originating from a smooth coarse-graining, the Fourier-space Heaviside window function and the Gaussian window function are both considered in what follows. 

With the power spectrum (\ref{eq:psG}), the covariance of $\delta (\tau)$ is given by 
\begin{subequations}
	\begin{align}
		\expval{ \delta (\tau_{1}) \delta (\tau_{2}) } 
		&= \frac{16}{81} 
		\frac{\mathcal{P}_{0}}{ \sqrt{2 \pi} } \, \frac{1}{\widetilde{k}_{\star}^{\alpha - 1}} 
		\frac{1}{ \widetilde{\tau}_{1}^{2} \widetilde{\tau}_{2}^{2} } 
		\int_{0}^{\infty} \frac{\dd x}{x} \, x^{\alpha + 4} \exp \qty[ - \frac{ (x - \widetilde{k}_{\star} )^{2} }{2} ] 
		\widetilde{\mathrm{W}} (z_{1}) \widetilde{\mathrm{W}} (z_{2}) 
		\notag \\ 
		&= \frac{16}{81} 
		\frac{\mathcal{P}_{0}}{ \sqrt{2 \pi} } \, \frac{1}{\widetilde{k}_{\star}^{\alpha - 1}} 
		\frac{\widetilde{\tau}_{1} \widetilde{\tau}_{2}}{4} I_{1} ( \widetilde{\tau}_{1}, \, \widetilde{\tau}_{2}) 
		\,\, .
	\end{align}
\end{subequations}

\subsubsection{Heaviside window function}

As in Section~\ref{sec:stexs}, let us start with the Fourier-space Heaviside window function as our benchmark. 
The discussion of Ref.~\cite{Kushwaha:2025zpz} (in which $\alpha = 0$) is essentially reviewed, as well as it is emphasised here that the uncoloured noise dominates first and then the coloured noise plays a relevant role. 
The considered window function is fixed to be  
\begin{equation}
	\widetilde{\mathrm{W}} (z) 
	= \Theta (1 - z) 
	\,\, . 
	\label{eq:ncov_wfH}
\end{equation}
In addition to Eq.~(\ref{eq:ncov_wfH}), $\alpha = 2$ is assumed hereafter, although the integrals appearing in Eqs.~(\ref{eq:ncovG_igen}) can be performed analytically for a general $\alpha$. 
This is because, when the Gaussian window function is considered in Section~\ref{subsec:ncov_g}, the variance of $\delta$ may have a finite nonzero value at $\tau = 0$, especially for $\alpha = 0$. 
This motivates us to use a nonzero $\alpha$, and as a demonstration $\alpha = 2$ here, to avoid such an undesirable behaviour. 
With this choice, the difference purely originating from the choice of the window function can be extracted in Section~\ref{sec:cnmf}. 

With Eq.~(\ref{eq:ncov_wfH}), one obtains 
\begin{subequations}
	\label{eq:ncovG_iH}
	\begin{align}
		I_{1} (\widetilde{\tau}_{1}, \, \widetilde{\tau}_{2}) 
		&= \frac{4}{ \widetilde{\tau}_{1}^{3} \widetilde{\tau}_{2}^{3} } 
		\left\{ 
			( \widetilde{k}_{\star}^{4} + 9 \widetilde{k}_{\star}^{2} + 8 ) 
			\exp \bigg( - \frac{ \widetilde{k}_{\star}^{2} }{2} \bigg) 
		\right. 
		\notag \\ 
		&\quad - [ \widetilde{k}_{\star}^{4} + \widetilde{k}_{\star}^{3} \widetilde{\tau}_{\wedge} + \widetilde{k}_{\star}^{2} ( \widetilde{\tau}_{\wedge}^{2} + 9 ) + \widetilde{k}_{\star} ( \widetilde{\tau}_{\wedge}^{3} + 7 \widetilde{\tau}_{\wedge} ) + ( \widetilde{\tau}_{\wedge}^{4} + 4 \widetilde{\tau}_{\wedge}^{2} + 8 ) ] 
		\exp \qty[ - \frac{ ( \widetilde{\tau}_{\wedge} - \widetilde{k}_{\star} )^{2} }{2} ] 
		\notag \\ 
		&\quad \left. 
			+ \, \widetilde{k}_{\star} ( \widetilde{k}_{\star}^{4} + 10 \widetilde{k}_{\star}^{2} + 15 ) 
			\cdot \sqrt{ \frac{\pi}{2} } \, \qty[ 
				\mathrm{erf} \bigg( \frac{ \widetilde{k}_{\star} }{\sqrt{2}} \bigg) 
				+ \mathrm{erf} \bigg( \frac{ \widetilde{\tau}_{\wedge} - \widetilde{k}_{\star} }{ \sqrt{2} } \bigg) 
			] 
		\right\} 
		\,\, , 
		\label{eq:ncovG_i1H}
		\\ 
		I_{2} (\widetilde{\tau}_{1}, \, \widetilde{\tau}_{2}) + I_{2} (\widetilde{\tau}_{2}, \, \widetilde{\tau}_{1}) 
		&= 2 \bigg( \frac{\widetilde{\tau}_{\wedge}}{\widetilde{\tau}_{\vee}} \bigg)^{3}  
		\exp \qty[ - \frac{ (\widetilde{\tau}_{\wedge} - \widetilde{k}_{\star})^{2} }{2} ] 
		\,\, , 
		\label{eq:ncovG_i2H}
		\\ 
		I_{3} (\widetilde{\tau}_{1}, \, \widetilde{\tau}_{2}) 
		&= \widetilde{\tau}_{1} \exp \qty[ - \frac{ ( \widetilde{\tau}_{1} - \widetilde{k}_{\star} )^{2} }{2} ] 
		\delta_{\rm D} ( \widetilde{\tau}_{1} - \widetilde{\tau}_{2} ) 
		\,\, . 
		\label{eq:ncovG_i3H}
	\end{align}
\end{subequations}
In Eqs.~(\ref{eq:ncovG_iH}), we have introduced 
\begin{equation*}
	\widetilde{\tau}_{\wedge} 
    \equiv \mathrm{min} (\widetilde{\tau}_{1}, \, \widetilde{\tau}_{2}) 
	\,\, , 
	\qquad 
	\widetilde{\tau}_{\vee} 
    \equiv \mathrm{max} (\widetilde{\tau}_{1}, \, \widetilde{\tau}_{2}) 
	\,\, . 
\end{equation*}
One notices that in $I_{3}$ the Dirac $\delta$-function appears, so the function $I_{3}$ plays the role of uncoloured noise. 
On the other hand, $I_{1}$ and $I_{2}$ are the coloured noises. 
The covariance matrix of the noise is then given by 
\begin{equation}
	\Delta^{2} \expval{ \xi (\tau_{1}) \xi (\tau_{2}) } 
	= \frac{16}{81} \frac{\mathcal{P}_{0}}{ \sqrt{2 \pi} } \frac{1}{\widetilde{k}_{\star}} 
	\qty[ 
		\text{Eq.~(\ref{eq:ncovG_i1H})} - \text{Eq.~(\ref{eq:ncovG_i2H})} + \text{Eq.~(\ref{eq:ncovG_i3H})} 
	] 
	\,\, . 
	\label{eq:ncovH_ms}
\end{equation}

\begin{figure}
	\centering
	\includegraphics[width = 0.995\linewidth]{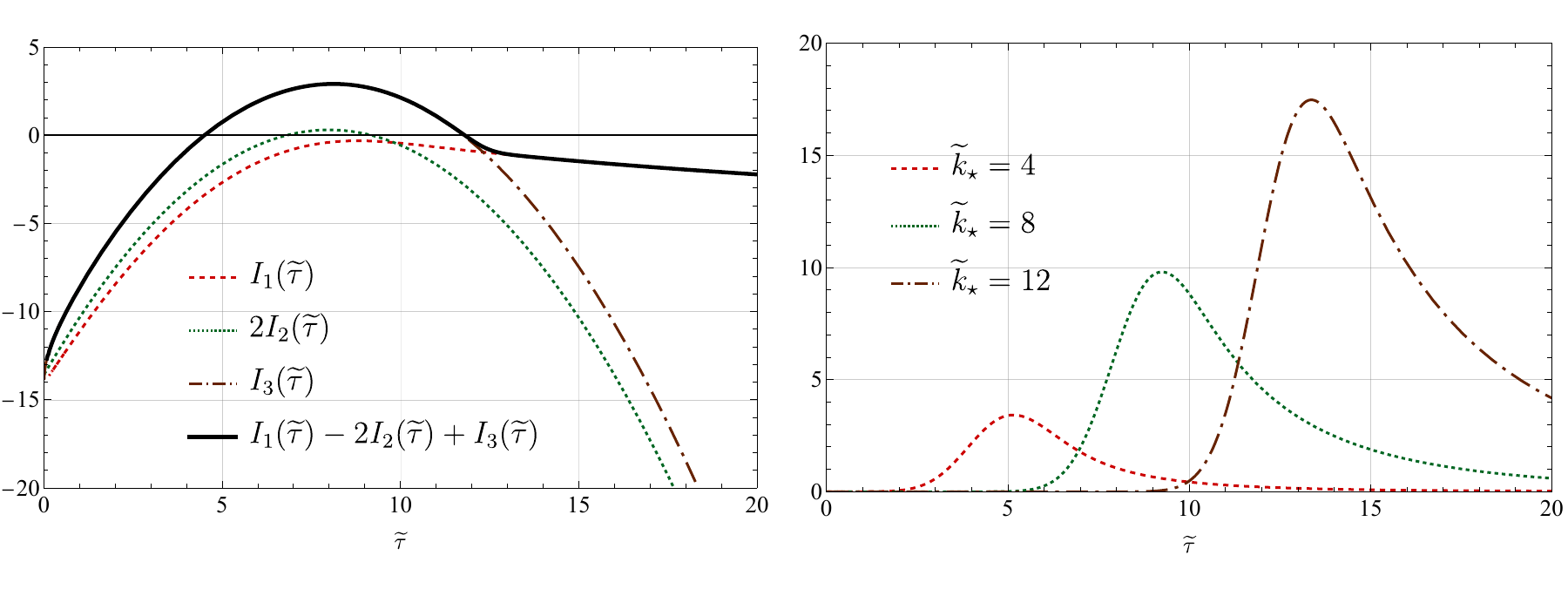}
	\caption{
		(\textit{Left}) 
		The three functions in the equal-time limit on a base-$10$ logarithmic scale, appearing in the variance of the noise (\ref{eq:ncovH_ms}) for the Heaviside window function and for $\widetilde{k}_{\star} = 8$. 
		(\textit{Right}) 
		The function $\widetilde{\tau}^{2} I_{1} ( \widetilde{\tau} ) / 4$, proportional to the variance of the coarse-grained density contrast for several $\widetilde{k}_{\star}$. 
		Contrary to the standard excursion-set method, the variance is no longer a monotonic function. 	} 
	\label{fig:ncov_varH}
\end{figure}

Let us focus on the $\widetilde{\tau}_{1} = \widetilde{\tau}_{2}$ limit to see the relevance of each component. 
Define 
\begin{equation}
	I_{1} (\widetilde{\tau}) 
	\equiv \lim_{\widetilde{\tau}_{2} \to \widetilde{\tau}} I_{1} (\widetilde{\tau}, \, \widetilde{\tau}_{2}) 
	\,\, , 
	\qquad 
	I_{2} (\widetilde{\tau}) 
	\equiv \lim_{\widetilde{\tau}_{2} \to \widetilde{\tau}} I_{2} (\widetilde{\tau}, \, \widetilde{\tau}_{2}) 
	\,\, , 
	\qquad 
	I_{3} (\widetilde{\tau}) 
	\equiv \lim_{\widetilde{\tau}_{2} \to \widetilde{\tau}} I_{3} (\widetilde{\tau}, \, \widetilde{\tau}_{2}) 
	\,\, . 
	\label{eq:gen_eqt}
\end{equation}
The limit can be evaluated straightforwardly for $I_{1}$ and $I_{2}$. 
The limit for Eq.~(\ref{eq:ncovG_i3H}) can in practice be considered by replacing the $\delta$-function with its discretised form, \textit{i.e.}~$\delta_{\rm D} \qty(\widetilde{\tau}_{i} - \widetilde{\tau}_{j}) \to \delta_{ij} / \Delta \widetilde{\tau}$, where $\Delta \widetilde{\tau} = 0.01$ was implemented in our numerical simulations. 
The left panel in Figure~\ref{fig:ncov_varH} shows those three functions on a base-$10$ logarithmic scale, in which one finds that the contribution from the uncoloured noise, the function $I_{3}$, dominates over the others up to a certain scale (slightly smaller scale than the fiducial scale $\widetilde{k}_{\star}$). 
The contribution from the coloured noise, especially the function $I_{1}$, starts to dominate later, which is because the localised power spectrum (\ref{eq:psG}) goes to zero on smaller scales, and so does the variance of the coarse-grained density contrast. 
Those observations will explain the stochastic trajectories numerically generated in Section~\ref{subsec:cnmf_esc}, which are also related to the behaviour of the resultant mass function of PBHs in Section~\ref{subsec:cnmf_mf}, where, contrary to the case with a smooth window function, both the probabilities $P_{1}$ and $P_{2}$ almost equally contribute to the relevant abundance of PBHs. 

The right panel in Figure~\ref{fig:ncov_varH} shows the function $\widetilde{\tau}^{2} I_{1} (\widetilde{\tau}) / 4$ for several $\widetilde{k}_{\star}$. 
It is proportional to the variance of the coarse-grained density contrast, 
\begin{equation}
	\expval{ [ \delta (\tau) ]^{2} } 
	= \frac{16}{81} \frac{\mathcal{P}_{0}}{ \sqrt{2 \pi} \, \widetilde{k}_{\star} } 
	\frac{ \widetilde{\tau}^{2} I_{1} (\widetilde{\tau}) }{4} 
	\,\, . 
	\label{eq:gen_I1d}
\end{equation}
When $\tau \to 0$, for $I_{1}$, the quantities in the curly brackets in Eq.~(\ref{eq:ncovG_i1H}) non-trivially cancel up to the fifth order in $\tau$, and the first nonvanishing term is thus proportional to $\tau^{6}$, giving rise to $I_{1} (\widetilde{\tau} \to 0) = (2/3) \exp \, ( - \widetilde{k}_{\star}^{2} / 2 )$ and, in passing, $I_{2} (\widetilde{\tau} \to 0) = 2 \exp \, ( - \widetilde{k}_{\star}^{2} / 2 )$. 
Given that $I_{1} (\widetilde{\tau} = 0) > 0$, the condition that $\expval{ [ \delta (\tau = 0) ]^{2} } = 0$
is automatically guaranteed. 

\subsubsection{Gaussian window function}
\label{subsec:ncov_g}

Next, let us consider the Gaussian window function as an example of a smooth coarse-graining, 
\begin{equation}
	\widetilde{\mathrm{W}} (z) 
	= \exp \qty( - \frac{z^{2}}{2} ) 
	\,\, . 
	\label{eq:wfG}
\end{equation}
For Eq.~(\ref{eq:wfG}), the functions defined in Eqs.~(\ref{eq:ncovG_igen}) are given by 
\begin{subequations}
	\label{eq:ncovG_iG}
	\begin{align}
		I_{1} (\widetilde{\tau}_{1}, \, \widetilde{\tau}_{2}) 
		&= 4 
		\exp \bigg( 
        			- \frac{ \widetilde{k}_{\star}^{2} }{2} 
    		\bigg) 
    		\qty( 
        			\frac{ \widetilde{\tau}_{1} \widetilde{\tau}_{2} }{ \widetilde{\tau}_{1}^{2} + \widetilde{\tau}_{1}^{2} \widetilde{\tau}_{2}^{2} + \widetilde{\tau}_{2}^{2} } 
		)^{5} 
    		\widetilde{\tau}_{1}^{2} \widetilde{\tau}_{2}^{2} 
    		\left\{ 
    			\vphantom{
            			\qty[ 
                				1 + \mathrm{erf} \bigg( 
                    				\frac{ \widetilde{k}_{\star} }{ \sqrt{2}   \, f } 
                				\bigg) 
            			] 
        			} 
    			( \widetilde{k}_{\star}^{4} + 9 \widetilde{k}_{\star}^{2} f^{2} + 8 f^{4} ) 
    		\right. 
    		\notag \\ 
    		&\quad + \left. 
        			\sqrt{\pi} \bigg( 
            			\frac{ \widetilde{k}_{\star} }{ \sqrt{2} \, f } 
        			\bigg) 
        			\exp \bigg( 
            			\frac{ \widetilde{k}_{\star}^{2} }{ 2 f^{2} } 
        			\bigg) 
        			\qty[ 
            			1 + \mathrm{erf} \bigg( 
                				\frac{ \widetilde{k}_{\star} }{ \sqrt{2} \, f } 
            			\bigg) 
        			] 
        			( \widetilde{k}_{\star}^{4} + 10 \widetilde{k}_{\star}^{2} f^{2} + 15 f^{4} ) 
		\right\} 
		\,\, , 
		\label{eq:ncovG_i1G}
		\\ 
		I_{2} (\widetilde{\tau}_{1}, \, \widetilde{\tau}_{2})  
		&= 
    	2 \exp \bigg( 
        			- \frac{ \widetilde{k}_{\star}^{2} }{2} 
		\bigg) 
    		\qty( 
        			\frac{ \widetilde{\tau}_{1} \widetilde{\tau}_{2} }{ \widetilde{\tau}_{1}^{2} + \widetilde{\tau}_{1}^{2} \widetilde{\tau}_{2}^{2} + \widetilde{\tau}_{2}^{2} } 
    		)^{7} 
            \widetilde{\tau}_{1}^{4}
            \widetilde{\tau}_{2}^{2} 
    		\left\{ 
			\vphantom{
            			\qty[ 
                				1 + \mathrm{erf} \bigg( 
                    				\frac{ \widetilde{k}_{\star} }{ \sqrt{2}   \, f } 
                				\bigg) 
            			] 
        			} 
			( \widetilde{k}_{\star}^{6} + 20 \widetilde{k}_{\star}^{4} f^{2} + 87 \widetilde{k}_{\star}^{2} f^{4} + 48 f^{6} ) 
		\right. 
		\notag \\[2.0ex] 
    		&\quad + \left. 
        			\sqrt{\pi} \bigg( 
            			\frac{ \widetilde{k}_{\star} }{ \sqrt{2} \, f } 
        			\bigg) 
        			\exp \bigg( 
            			\frac{ \widetilde{k}_{\star}^{2} }{ 2 f^{2} } 
        			\bigg) 
        			\qty[ 
            			1 + \mathrm{erf} \bigg( 
                				\frac{ \widetilde{k}_{\star} }{ \sqrt{2} \, f } 
            			\bigg) 
        			] 
        			( \widetilde{k}_{\star}^{6} + 21 \widetilde{k}_{\star}^{4} f^{2} + 105 \widetilde{k}_{\star}^{2} f^{4} + 105 f^{6} ) 
    		\right\} 
		\,\, , 
		\label{eq:ncovG_i2G}
		\\ 
		I_{3} (\widetilde{\tau}_{1}, \, \widetilde{\tau}_{2}) 
		&= \exp \bigg( 
        			- \frac{ \widetilde{k}_{\star}^{2} }{2} 
    		\bigg) 
    		\qty( 
        			\frac{ \widetilde{\tau}_{1} \widetilde{\tau}_{2} }{ \widetilde{\tau}_{1}^{2} + \widetilde{\tau}_{1}^{2} \widetilde{\tau}_{2}^{2} + \widetilde{\tau}_{2}^{2} } 
    		)^{9} 
    		\widetilde{\tau}_{1}^{4} \widetilde{\tau}_{2}^{4} 
		\left\{ 
			\vphantom{
            			\qty[ 
                				1 + \mathrm{erf} \bigg( 
                    				\frac{ \widetilde{k}_{\star} }{ \sqrt{2}   \, f } 
                				\bigg) 
            			] 
        			} 
			 ( \widetilde{k}_{\star}^{8} + 35 \widetilde{k}_{\star}^{6} f^{2} + 345 \widetilde{k}_{\star}^{4} f^{4} + 975 \widetilde{k}_{\star}^{2} f^{6} + 384 f^{8} ) 
		\right. 
    		\notag \\[2.0ex] 
    		&\quad + \left. 
        			\sqrt{\pi} \bigg( 
            			\frac{ \widetilde{k}_{\star} }{ \sqrt{2} \, f } 
        			\bigg) 
        			\exp \bigg( 
            			\frac{ \widetilde{k}_{\star}^{2} }{ 2 f^{2} } 
        			\bigg) 
        			\qty[ 
            			1 + \mathrm{erf} \bigg( 
                				\frac{ \widetilde{k}_{\star} }{ \sqrt{2} \, f } 
            			\bigg) 
        			] 
        			\qty( \widetilde{k}_{\star}^{8} + 36 \widetilde{k}_{\star}^{6} f^{2} + 378 \widetilde{k}_{\star}^{4} f^{4} + 1260 \widetilde{k}_{\star}^{2} f^{6} + 945 f^{8} ) 
    		\right\} 
    		\,\, . 
		\label{eq:ncovG_i3G}
	\end{align}
\end{subequations}
Here, $f$ is a function that depends on the time variable, defined by 
\begin{equation}
	f 
	= f \qty(\widetilde{\tau}_{1}, \, \widetilde{\tau}_{2}) 
	\equiv \sqrt{ 
		1 + \frac{1}{ \widetilde{\tau}_{1}^{2} } + \frac{1}{ \widetilde{\tau}_{2}^{2} } 
	} 
	\,\, . 
\end{equation}
For those functions, the covariance matrix of the noise is given by 
\begin{equation}
	\Delta^{2} \expval{ \xi (\tau_{1}) \xi (\tau_{2}) } 
	= \frac{16}{81} \frac{\mathcal{P}_{0}}{ \sqrt{2 \pi} } \frac{1}{\widetilde{k}_{\star}} 
	\qty{ 
		\text{Eq.~(\ref{eq:ncovG_i1G})} - 
        \qty[ 
            \text{Eq.~(\ref{eq:ncovG_i2G})} + \eval{ 
                \text{Eq.~(\ref{eq:ncovG_i2G})} 
            }_{\widetilde{\tau}_{1} \leftrightarrow \widetilde{\tau}_{2}} 
        ] 
        + \text{Eq.~(\ref{eq:ncovG_i3G})} 
    } 
	\,\, . 
	\label{eq:ncovG_ms}
\end{equation}

\begin{figure}[h]
	\centering
	\begin{minipage}[b]{0.495\linewidth}
		\centering
		\includegraphics[width=0.995\linewidth]{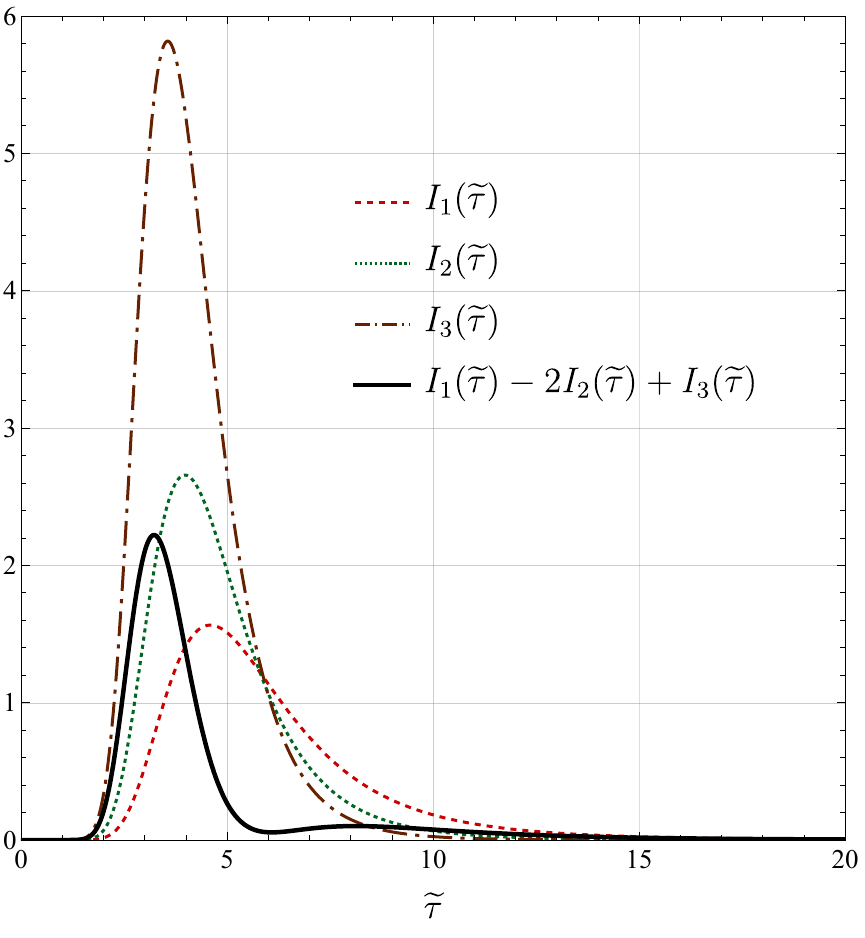}
	\end{minipage}
	\begin{minipage}[b]{0.495\linewidth}
		\centering
		\includegraphics[width=0.995\linewidth]{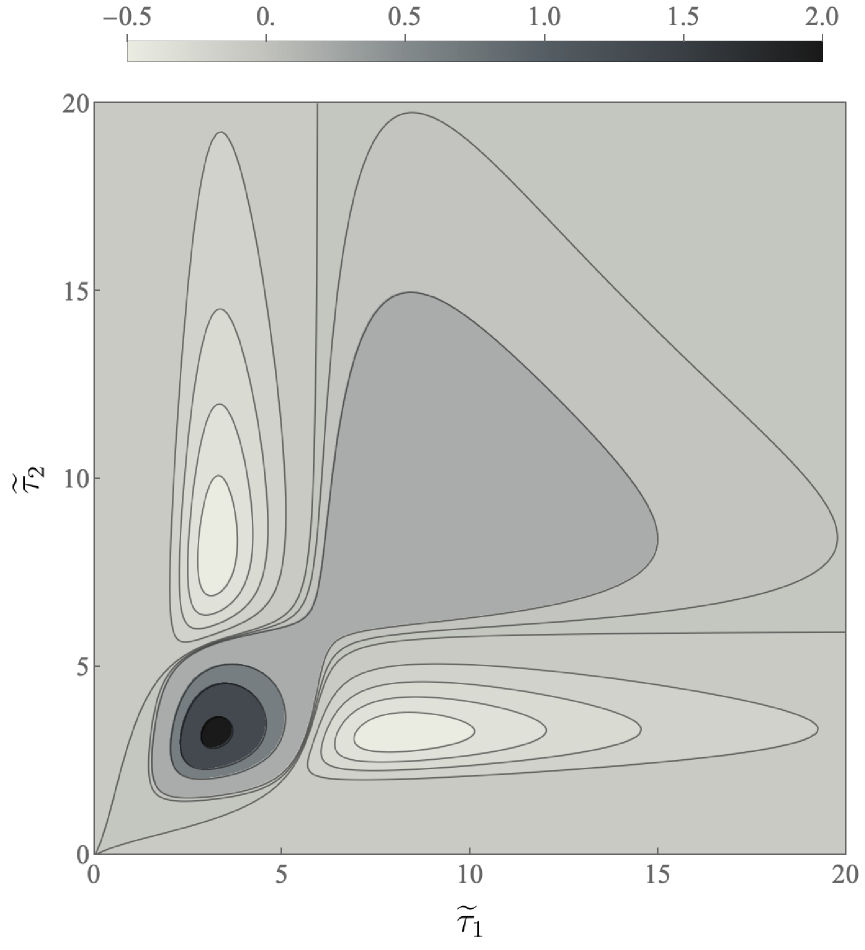}
	\end{minipage}
	\caption{
		(\textit{Left}) The diagonal elements of the functions appearing in the covariance of the noise, the limit $\widetilde{\tau}_{1} = \widetilde{\tau}_{2} = \widetilde{\tau}$ in Eq.~(\ref{eq:ncovG_ms}). 
		(\textit{Right}) The combination 
        $I_{1}(\widetilde{\tau}_{1}, \widetilde{\tau}_{2}) - [ I_{2} (\widetilde{\tau}_{1}, \widetilde{\tau}_{2}) + I_{2} (\widetilde{\tau}_{2}, \widetilde{\tau}_{1}) ] + I_{3}(\widetilde{\tau}_{1}, \widetilde{\tau}_{2})$ 
        including the non-diagonal components, Eq.~(\ref{eq:ncovG_ms}). 
		The parameters are fixed to be $k_{\star} = 8$ 
        in both panels. 
	}
	\label{fig:noisecov_g}
\end{figure}

The left panel in Figure~\ref{fig:noisecov_g} shows the diagonal element of the functions, $I_{1}$, $I_{2}$, and $I_{3}$, defined in Eq.~(\ref{eq:gen_eqt}), and their combination appearing in Eq.~(\ref{eq:ncovG_ms}). 
The parameters are fixed to be $\widetilde{k}_{\star} = 8$ and $\mathcal{P}_{0} = 1$. 
The difference from the previous case is that every contribution is coloured, and that all three functions non-negligibly contribute irrespective of the value of $\tau$. 
The right panel in Figure~\ref{fig:noisecov_g} shows the contour plot of the function $I_{1} (\widetilde{\tau}_{1}, \, \widetilde{\tau}_{2}) - [ I_{2} (\widetilde{\tau}_{1}, \, \widetilde{\tau}_{2}) + I_{2} (\widetilde{\tau}_{2}, \, \widetilde{\tau}_{1}) ] + I_{3} (\widetilde{\tau}_{1}, \, \widetilde{\tau}_{2})$, which, together with the left panel, will be used for consistency check when the coloured noise is numerically generated in Section~\ref{sec:cnmf}. 

When $\tau_{1} = \tau_{2} = \tau \to 0$, the function $f$ behaves as $f \approx \sqrt{2} / \widetilde{\tau}$. 
The very last factors in the curly braces in Eqs.~(\ref{eq:ncovG_iG}) therefore become the most dominant contribution. 
In this limit, one obtains 
\begin{equation}
	I_{1} (\widetilde{\tau}) 
	= 4 \exp \bigg( - \frac{ \widetilde{k}_{\star}^{2} }{2} \bigg) 
	\,\, , 
	\qquad 
	I_{2} (\widetilde{\tau}) 
	= 6 \exp \bigg( - \frac{ \widetilde{k}_{\star}^{2} }{2} \bigg) 
	\,\, , 
	\qquad 
	I_{3} (\widetilde{\tau}) 
	= 12 \exp \bigg( - \frac{ \widetilde{k}_{\star}^{2} }{2} \bigg) 
	\,\, , 
\end{equation}
so that $- 2 I_{2} (\widetilde{\tau})$ and $I_{3} (\widetilde{\tau})$ cancel each other. 
One also sees that the requirement for the coarse-grained density contrast in $\tau \to 0$ limit, $\expval{ [ \delta (\tau) ]^{2} } \to 0$ 
for Eq.~(\ref{eq:gen_I1d}), is again properly satisfied. 
This ceases to hold when, for instance, $\alpha = 0$, which explains why the power-law factor $(k / k_{\star})^{\alpha}$ is introduced in Eq.~(\ref{eq:psG}). 

Now that all the relevant analytical formulas for the noise covariance have been obtained. 
The noises with those correlation properties can be numerically generated, which can be used to simulate the stochastic process exhibited by the coarse-grained density contrast. 
From the fraction that the stochastic realisations cross the threshold, required to form a PBH, the mass function can then be numerically reconstructed. 

\section{Coloured noise effects on mass function}
\label{sec:cnmf}

Though there are several numerical ways to simulate the stochastic process driven by the coloured noise, generation of coloured noises with the desired correlation properties is far from trivial. 
A convenient method relies on the Cholesky factorisation, sometimes also called the Cholesky decomposition, which factorises a positive-definite matrix such as the covariance matrix into the product of the lower-triangular and upper-triangular matrices. 
The scheme is explained in Section~\ref{subsec:cnmf_cfs}, with a concrete demonstration for the Ornstein--Uhlenbeck process, an analytically solvable stochastic process that generates the coloured noises. 
Section~\ref{subsec:cnmf_esc} then numerically solves the stochastic process driven by the coloured noises, analytically discussed in the previous section, as well as the consistency of the generated noises is confirmed. 
The mass function of primordial black holes resulting from stochastic processes driven by the coloured noises is numerically derived in Section~\ref{subsec:cnmf_mf}. 

\subsection{Cholesky factorisation}
\label{subsec:cnmf_cfs}

A $d \times d$ positive-definite matrix $\Sigma$ can in general be factorised into the lower-triangular matrix $L$ and its transpose $L^{\mathsf{T}}$, as 
\begin{equation}
    \Sigma = L L^{\mathsf{T}} 
    \,\, . 
\end{equation}
Suppose that a vector $\vb*{x} \sim N (\vb*{0}, \, 1_{d \times d})$ is prepared, in which each element $x_{i}$ is a random number generated from the Gaussian distribution with zero mean and unit variance ($1_{d \times d}$ denotes the unit matrix of dimension $d$). 
The product $\vb*{y} = L \vb*{x}$ then generates the coloured noise with the covariance $\Sigma$. 
Indeed, 
\begin{equation}
    \expval{ \vb*{y} \vb*{y}^{\mathsf{T}} } 
    = \expval{ L \vb*{x} \vb*{x}^{\mathsf{T}} L^{\mathsf{T}} } 
    = L \expval{ \vb*{x} \vb*{x}^{\mathsf{T}} } L^{\mathsf{T}} 
    = L 1_{d \times d} L^{\mathsf{T}} 
    = L L^{\mathsf{T}} 
    = \Sigma 
    \,\, . 
\end{equation}
The matrix $\Sigma$ will later be identified with the nondimensionalised and discretised version of the covariance matrix, $\Sigma_{ij} = \Delta^{2} \expval{ \xi (\tau_{i}) \xi (\tau_{j}) }$, in our numerical simulation. 
In this way the correlated noise $\xi (\tau)$ itself at each time step can be generated, which drives the stochastic process exhibited by the coarse-grained density contrast. 

\paragraph*{\textbf{\textit{Ornstein--Uhlenbeck process.}}}

Let us demonstrate here the above procedure for the coloured noise generated by an analytically solvable stochastic process. 
The stochastic process described by 
\begin{equation}
    \dv{ \xi (t) }{ t }  
    = - a \xi (t) + b \, \dv{W (t)}{t} 
    \label{eq:ouproc}
\end{equation}
is called the Ornstein--Uhlenbeck process. 
Here, $a > 0$ and $b > 0$ are constant, and $W(t)$ is a Wiener process so that $\dd W (t) / \dd t$ is an uncoloured noise. 
The stochastic process (\ref{eq:ouproc}) can be solved by several ways to derive the covariance and higher moments of $\xi (t)$. 
For instance, one can solve the corresponding Fokker--Planck equation to first derive the distribution function of the stochastic variable $\xi (t)$. 
For the potential $V (\xi) \equiv a \xi^{2} / 2$, it reads 
\begin{equation}
	\pdv{ f }{ t } 
	= \pdv{ \xi } \qty( \dv{V}{\xi} f ) + \frac{b^{2}}{2} \pdv[2]{f}{\xi} 
	\,\, , 
	\qquad 
	f = f (\xi, \, t) = f (\xi , \, t \mid \xi_{0}, \, t_{0}) 
	\,\, . 
	\label{eq:ou_df}
\end{equation}
The solution to Eq.~(\ref{eq:ou_df}) can be obtained by the method of characteristics through the characteristic function, 
\begin{equation}
	\widetilde{f} (\eta, \, t \mid \xi_{0}, \, t_{0}) 
	\equiv \int \dd \xi \, e^{i \xi \eta} \, f (\xi, \, t \mid \xi_{0}, \, t_{0}) 
	\,\, . 
	\label{eq:chara}
\end{equation}
Then, Eq.~(\ref{eq:ou_df}) reduces to 
\begin{equation}
	\pdv{ \widetilde{f} }{ t } 
	+ a \eta \pdv{ \widetilde{f} }{ \eta } 
	= - \frac{b^{2} \eta^{2}}{2} \widetilde{f} 
	\,\, , 
	\qquad 
	\widetilde{f} 
	= \widetilde{f} (\eta, \, t) 
	= \widetilde{f} (\eta, \, t \mid \xi_{0}, \, t_{0}) 
	\,\, , 
\end{equation}
which can be solved exactly. 
The Lagrange--Charpit equation reads $\dd t = \dd \eta / a \eta = \dd \widetilde{f} / [ (- b^{2} \eta^{2} / 2) \widetilde{f} ]$, from which, together with the concentrated initial condition, \textit{i.e.}~$f (\xi, \, t = t_{0} \mid \xi_{0}, \, t_{0}) = \delta_{\rm D} (\xi - \xi_{0})$, one obtains $\widetilde{f} (\eta, \, t \mid \xi_{0}, \, t_{0}) = \exp \qty{ i \xi_{0} \eta e^{- a (t - t_{0})} - b^{2} \eta^{2} [ 1 - e^{- 2 a (t - t_{0})} ] / 4 a }$.
Substitution into Eq.~(\ref{eq:chara}) then gives the exact solution to Eq.~(\ref{eq:ou_df}), 
\begin{equation}
	f (\xi, \, t \mid \xi_{0}, \, t_{0}) 
	= \sqrt{ \frac{
    a}{\pi b^{2} \qty[ 1 - e^{- 2 a (t - t_{0})} ]} } \, \exp \qty{ 
		- \frac{ a }{ b^{2} } 
		\frac{ \qty[ \xi - \xi_{0} e^{- a (t - t_{0})} ]^{2} }{ 1 - e^{- 2 a (t - t_{0})} } 
	} 
	\,\, . 
\end{equation}

The covariance of $\xi (t)$ follows from Markov property of the stochastic process (\ref{eq:ouproc}) and law of total expectation that \begin{align}
	\Sigma (t_{1}, \, t_{2}) 
	&= \expval{ \xi (t_{1}) \xi (t_{2}) } 
	= \expval{ \xi (t_{\wedge}) \expval{ \xi (t_{\vee}) \mid \xi (t_{\wedge}) } } 
	= \expval{ [ \xi (t_{\wedge}) ]^{2} } e^{ - a (t_{\vee} - t_{\wedge}) } 
	\notag \\ 
	&= \frac{ b^{2} }{  2a } \qty[ 
		e^{  - a  (t_{\vee} - t_{\wedge}) } - e^{  - a  (t_1 + t_2) }
	] 
	= \frac{ b^{2} }{ 2 a }  \times 
	\begin{cases}
		e^{  - a  (t_{2} - t_{1})  } - e^{  - a  (t_{2} + t_{1})  } \,\, , &\quad t_{1} \leq t_{2} \,\, , \\[2.0ex] 
		e^{  - a  (t_{1} - t_{2})  } - e^{  - a  (t_{1} + t_{2})  } \,\, , &\quad t_{1} \geq t_{2} \,\, . 
	\end{cases}
    \label{eq:oucov}
\end{align}
For $\Sigma_{ij} = \Sigma (t_{i}, \, t_{j})$, time is discretised into $t_{k} = t_{0} + k \Delta t$, where the offset is introduced by $t_{0} = \Delta t$ and the indices run over $k = 0, \, 1, \, \dots, \, n - 1$. 
Let the lower-triangular matrix $L$ such that $\Sigma = L L^{\mathsf{T}}$ be 
\begin{equation}
    L 
    = \mqty( 
        L_{0, \, 0} & 0 & \cdots & 0 & 0  
        \\ 
        L_{1, \, 0} & L_{1, \, 1} & \cdots & 0 & 0 
        \\ 
        \vdots & \vdots & \ddots & \vdots & \vdots 
        \\ 
        L_{n-2, \, 0} & L_{n-2, \, 1} & \cdots & L_{n-2, \, n-2} & 0 
        \\ 
        L_{n-1, \, 0} & L_{n-1, \, 1} & \cdots & L_{n-1, \, n-2} & L_{n-1, \, n-1}
    ) 
    \,\, , 
    \qquad 
    L^{\mathsf{T}} 
    = \mqty( 
        L_{0, \, 0} & L_{1, \, 0} & \cdots & L_{n-2, \, 0} & L_{n-1, \, 0}  
        \\ 
        0 & L_{1, \, 1} & \cdots & L_{n-2, \, 1} & L_{n-1, \, 1} 
        \\ 
        \vdots & \vdots & \ddots & \vdots & \vdots 
        \\ 
        0 & 0 & \cdots & L_{n-2, \, n-2} & L_{n-1, \, n-2} 
        \\ 
        0 & 0 & \cdots & 0 & L_{n-1, \, n-1}
    ) 
    \,\, . 
    \label{eq:ltaz}
\end{equation}
Each element in Eq.~(\ref{eq:ltaz}) can be determined by 
\begin{equation}
    \sum_{k = 0}^{\mathrm{min} (i, \, j)} L_{i, \, k} L_{j, \, k} 
    = \frac{b^{2}}{a} \exp [ - a \cdot \mathrm{max} (t_{i}, \, t_{j}) ] \sinh [ a \cdot \mathrm{min} (t_{i}, \, t_{j}) ] 
    \,\, . 
\end{equation}
It is sufficient to consider $i \geq j$, and then for each $0 \leq j \leq n-1$, one obtains the system of $n-j$ equations, which can be solved recursively in increasing order of $j \leq i \leq n-1$, to arrive at 
\begin{equation}
	L 
	= 
	\qty[ \frac{b^{2}}{a} \sinh (a \Delta t) ]^{1/2} \, \exp \qty( - \frac{a \Delta t}{2} ) 
	\mqty( 
		1 & 0 & 0 & \cdots & \cdots & 0 \\
		e^{- a \Delta t} & 1 & 0 & \cdots & \cdots & 0 \\ 
		e^{- 2a \Delta t} & e^{- a \Delta t} & 1 & \cdots & \cdots & 0 \\ 
		\vdots & \vdots & \vdots & \ddots & \vdots & \vdots \\ 
		e^{- (n-2) a \Delta t} & e^{- (n-3) a \Delta t} & e^{- (n-4) a \Delta t} & \cdots & 1 & 0 \\ 
		e^{- (n-1) a \Delta t} & e^{- (n-2) a \Delta t} & e^{- (n-3) a \Delta t} & \cdots & e^{- a \Delta t} & 1 
	) 
	\,\, . 
\end{equation}
When the matrix $L$ is multiplied to a random vector $\vb*{x}$, the covariance of the resultant vector $\vb*{y} = L \vb*{x}$ respects Eq.~(\ref{eq:oucov}). 

\begin{figure}
	\centering
	\begin{minipage}[b]{0.495\linewidth}
		\centering
		\includegraphics[width=0.995\linewidth]{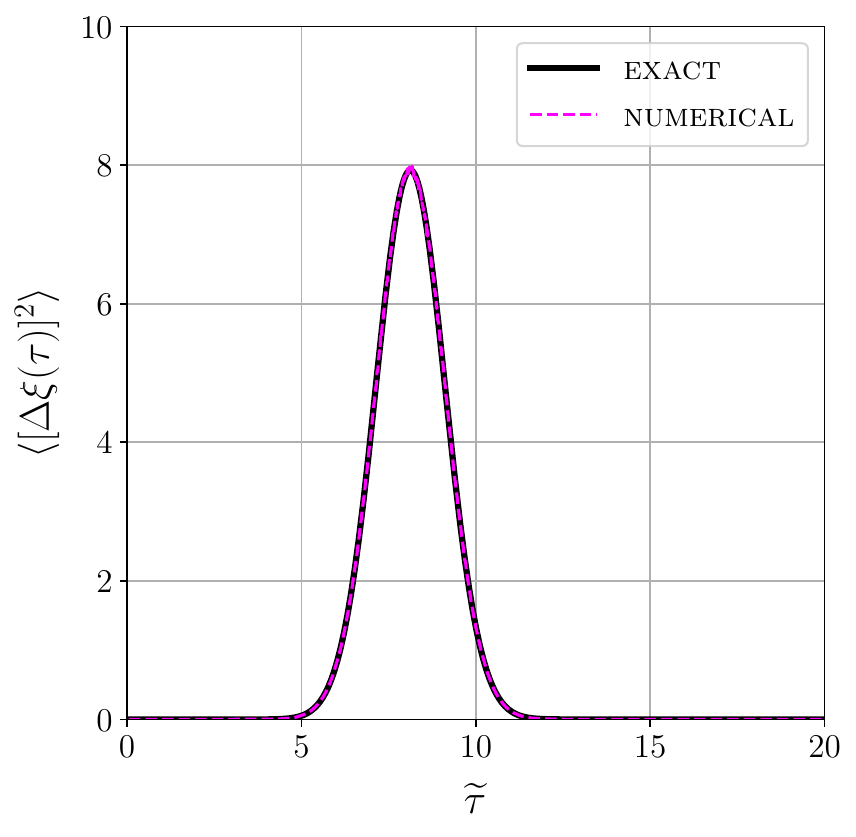}
	\end{minipage}
	\begin{minipage}[b]{0.495\linewidth}
		\centering 
        \includegraphics[width=0.995\linewidth]{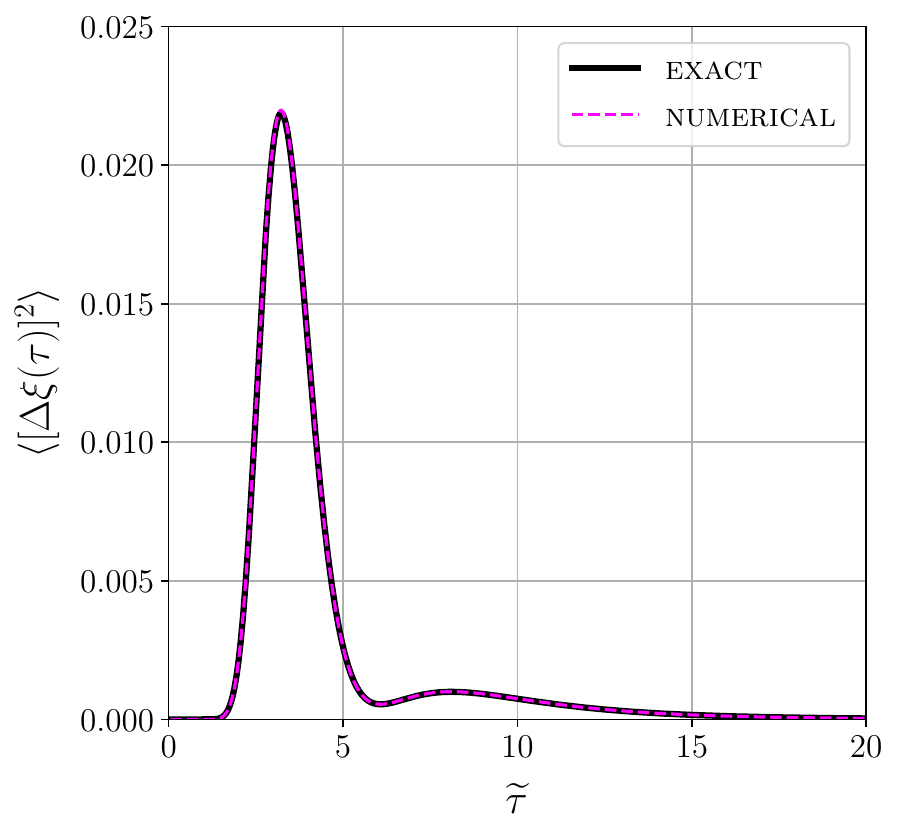}
	\end{minipage}
	\caption{
		The analytical and numerically reconstructed variance of the correlated noise, for $\widetilde{\mathrm{W}} (z) = \Theta (1 - z)$ and $\widetilde{\mathrm{W}} (z) = \exp \, (- z^{2} / 2)$ in the \textit{left} and \textit{right} panels, respectively. 
		The numerical curve was generated using the Cholesky factorisation with $N = 10^{5}$ noise realisations averaged. 
		The parameters are fixed to be $k_{\star} = 8$ and $\mathcal{P}_{0} = 1$ in both panels. 
	}
	\label{fig:noisecov_repr}
\end{figure}

\vspace{1em}

In practice, the Cholesky factorisation is performed numerically in most cases. 
In our context, the covariance matrices are analytically prepared in Eq.~(\ref{eq:ncovH_ms}) and Eq.~(\ref{eq:ncovG_ms}), based on which the lower-triangular matrix $L$ can be obtained numerically. 
The consistency of the Cholesky factorisation for the generation of the desired coloured noise has been confirmed in Figure~\ref{fig:noisecov_repr}, where the left and right panels are based on Eq.~(\ref{eq:ncovH_ms}) and Eq.~(\ref{eq:ncovG_ms}) respectively. 
The difference between those two panels is the window function used, while the fiducial scale divided by the width of the power spectrum is fixed at $\widetilde{k}_{\star} = k_{\star} / \Delta = 8$ for illustrative purposes. 
Though the figures are only for the variance, the covariance at different times can also be confirmed in the same way. 
Now, the stochastic process exhibited by $\delta (\tau)$ with respect to the variation of $\tau = 1 / R$ can therefore be numerically simulated. 
In those simulations, one vector $\vb*{y} = L \vb*{x}$ corresponds to one stochastic trajectory. 
That is, starting from the initial condition $\delta (\tau = 1/R = 0) = 0$, at each coarse-graining scale the coloured noise is injected to calculate the coarse-grained density contrast at a different (smaller) scale, to generate one stochastic trajectory. 
Therefore, to draw the statistical consequences, a sufficiently large number of the vectors $\vb*{x}$ and $\vb*{y} = L \vb*{x}$ should be prepared, and, in Figure~\ref{fig:noisecov_repr}, $N = 10^{5}$ stochastic realisations were generated and then averaged. 

\subsection{Excursion-set evolution driven by coloured noise}
\label{subsec:cnmf_esc}

\begin{figure}
	\centering
	\begin{minipage}[b]{0.495\linewidth}
		\centering
        \includegraphics[width=0.995\linewidth]{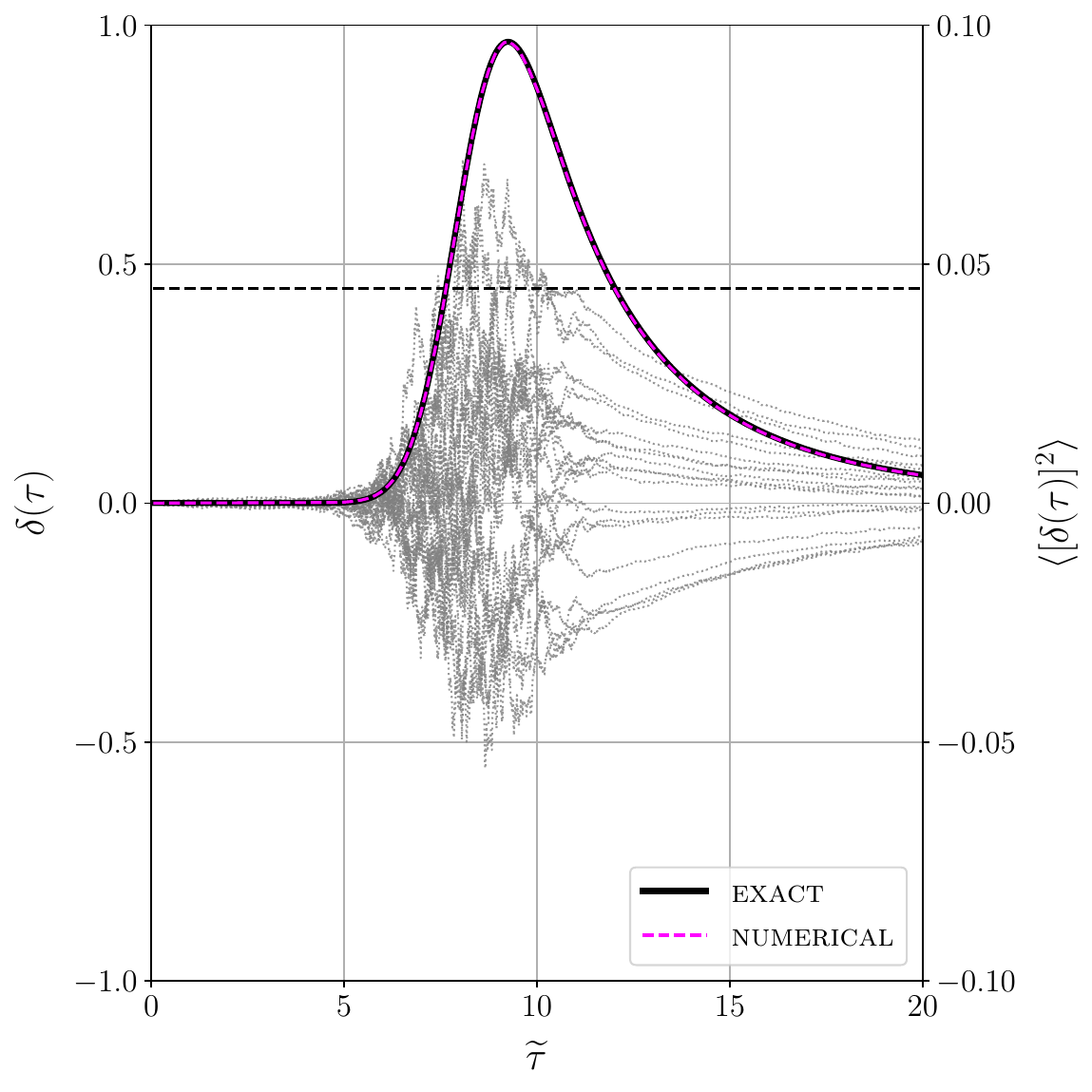}
	\end{minipage}
	\begin{minipage}[b]{0.495\linewidth}
		\centering 
        \includegraphics[width=0.995\linewidth]{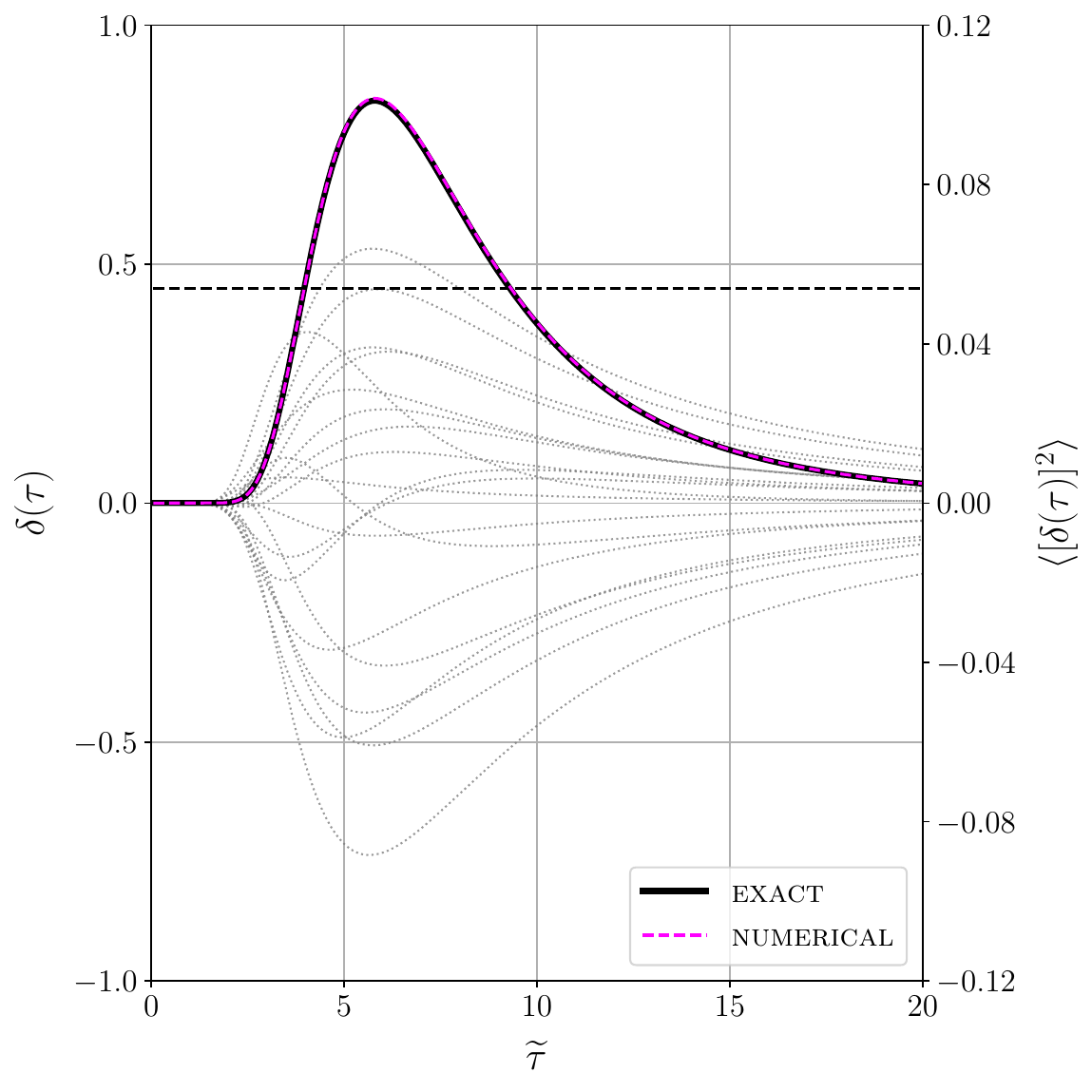}
	\end{minipage}
	\caption{
		Twenty samples of the stochastic trajectories for $\widetilde{\mathrm{W}} (z) = \Theta (1 - z)$ and $\widetilde{\mathrm{W}} (z) = \exp \, (- z^{2} / 2)$, in the \textit{left} and \textit{right} panels respectively, together with the variance (thick curves). 
		The parameters are fixed to be $k_{\star} = 8$ and $\mathcal{P}_{0} = 1$ in both panels. 
	}
	\label{fig:rw_hfil_traj}
\end{figure}

Based on the Cholesky factorisation, the correlated noises were generated to numerically solve the stochastic process $\dd \delta (\tau) / \dd \tau = \xi (\tau)$. 
The left panel in Figure~\ref{fig:rw_hfil_traj} shows $20$ sample trajectories, for the case where $\delta$ is coarse-grained by the Fourier-space Heaviside window function while
the Gaussian window function is used in the right panel. 
The dashed horizontal line indicates the threshold value $\delta_{\rm c} = 0.45$, which follows Ref.~\cite{Kushwaha:2025zpz}.\footnote{
	The difference of the threshold values originating from the choice of the window function~\cite{Young:2019osy} is not accounted for in this article. 
}
That is, when $\delta (\tau) > \delta_{\rm c}$ is realised at a monitored scale $\tau = 1/R$, one considers that a PBH of the corresponding mass is formed. 
It should be noted that, in addition to the traditional criterion $\delta (\tau) > \delta_{\rm c}$ that we use throughout, several other criteria have been proposed such as the one based on the so-called compaction function~\cite{Shibata:1999zs, Harada:2023ffo, Harada:2024trx} (see also \textit{e.g.}~\cite{Musco:2018rwt, Escriva:2019phb,Musco:2020jjb}). 

An immediate observation is that, due to the more correlated nature of the noises, the stochastic trajectory is smoother for $\widetilde{\mathrm{W}} (z) = \exp \, (- z^{2} / 2)$ than for $\widetilde{\mathrm{W}} (z) = \Theta (1 - z)$, with respect to the change of the coarse-graining scale. 
In particular, in the left panel, the contribution from the uncoloured noise is prominent during $\widetilde{\tau} \lesssim \widetilde{k}_{\star}$, after which that from the coloured noise pulls the trajectories back towards $\delta (\tau) = 0$.  
The former behaviour originates from the window function, while the latter from the localised feature of the power spectrum. 
Those are consistent with Figure~\ref{fig:ncov_varH}. 
In other words, when the Fourier-space Heaviside window function is used, the dominant driving force that enables $\delta (\tau)$ to pierce the threshold is the uncoloured noise, though there are both uncoloured and coloured contributions. 
Under the Gaussian window function, on the other hand, the coloured noises are prevailing throughout the stochastic process, \textit{i.e.}~on all coarse-graining scales. 

The consistency of our numerical simulations with the analytical formula (\ref{eq:gen_I1d}) has been confirmed by averaging $N = 10^{5}$ stochastic realisations at each time (on each coarse-graining scale), giving rise to the variance of $\delta (\tau)$, also shown in Figure~\ref{fig:rw_hfil_traj} by the thick curves. 
With those stochastic realisations, the relevant probabilities and their derivatives, the latter of which are related to the mass function of PBHs, are now ready to be extracted. 

\subsection{Mass function of primordial black holes}
\label{subsec:cnmf_mf}

Now that all the ingredients have been gathered, the formation probability as well as the mass function of PBHs can numerically be reconstructed. 
For the numerically generated stochastic realisations, each trajectory is analysed to see whether, for each $\tau$, it exceeds the threshold or not. 
This procedure numerically reconstructs the two kinds of probabilities that will be defined in Eq.~(\ref{eq:mf_def}). 
What one calls the mass function of PBHs will then be extracted from those probabilities by differentiating then with respect to $\tau$, \textit{i.e.}~the mass scale. 

\subsubsection{Probabilities}
\label{subsec:probs}

As was generalised in Eq.~(\ref{eq:smdc_fgenp}), the formation probability of PBHs is given by the sum of the two probabilities, 
\begin{equation}
	P (\tau) 
	= P_{1} [ \mathrm{A} (\tau) ] + P_{2} [ \mathrm{B} (\tau) ] 
	\,\, . 
	\label{eq:mf_def}
\end{equation}
Here, note that the time variable is $\tau = 1/R$ instead of the variance of $\delta (\tau)$, since the latter is no longer a monotonic function with respect to $R$. 
The two events $\mathrm{A}$ and $\mathrm{B}$ are defined correspondingly by (see \textit{e.g.}~Ref.~\cite{Kushwaha:2025zpz})
\begin{subequations}
	\begin{align}
		\mathrm{A} (\tau) 
		&\equiv \qty{ \delta \mid \delta (\tau) > \delta_{\rm th} } 
		\,\, , 
		\\ 
		\mathrm{B} (\tau) 
		&\equiv \qty{ \delta \mid \delta (\tau) < \delta_{\rm th} \cap {}^{\exists} \tau' < \tau ~\text{such that}~ \delta (\tau') > \delta_{\rm th} }
		\,\, . 
	\end{align}
\end{subequations}
The left-hand side in Eq.~(\ref{eq:mf_def}) gives the (total) probability that a PBH with its mass larger than $M$ is formed. 
Now, for a fixed $\tau$, if a stochastic realisation exceeds $\delta_{\rm th}$, it is counted as an event $\mathrm{A}$. 
If the value $\delta (\tau)$ is less than $\delta_{\rm th}$, but there exists a smaller $\tau' < \tau$ for which $\delta (\tau')$ exceeds the threshold, it is counted as an event $\mathrm{B}$. 
The latter implies that there exists a larger coarse-graining scale (heavier mass scale) for which the formation criterion is fulfilled. 
Even if $\delta (\tau) < \delta_{\rm th}$ at the mass scale $\tau$, a PBH is thus considered to be formed (at the reference location $\vb*{x}$) that encloses any smaller structures at that location, so $P_{2}$ also contributes to the total formation probability. 
Recall that $P_{1} = P_{2}$ if the noise is not coloured at all, which can no longer be realised in the context of the formation of PBHs. 

\begin{figure}
	\centering
	\begin{minipage}[b]{0.495\linewidth}
		\centering
		\includegraphics[width=0.995\linewidth]{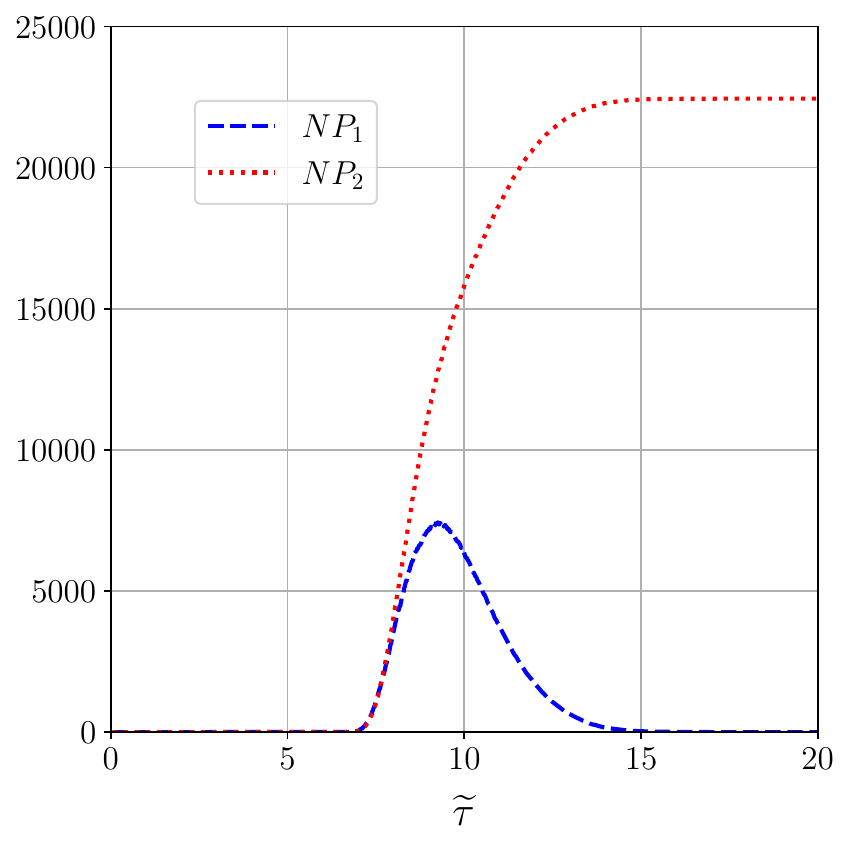}
	\end{minipage}
	\begin{minipage}[b]{0.495\linewidth}
		\centering        \includegraphics[width=0.995\linewidth]{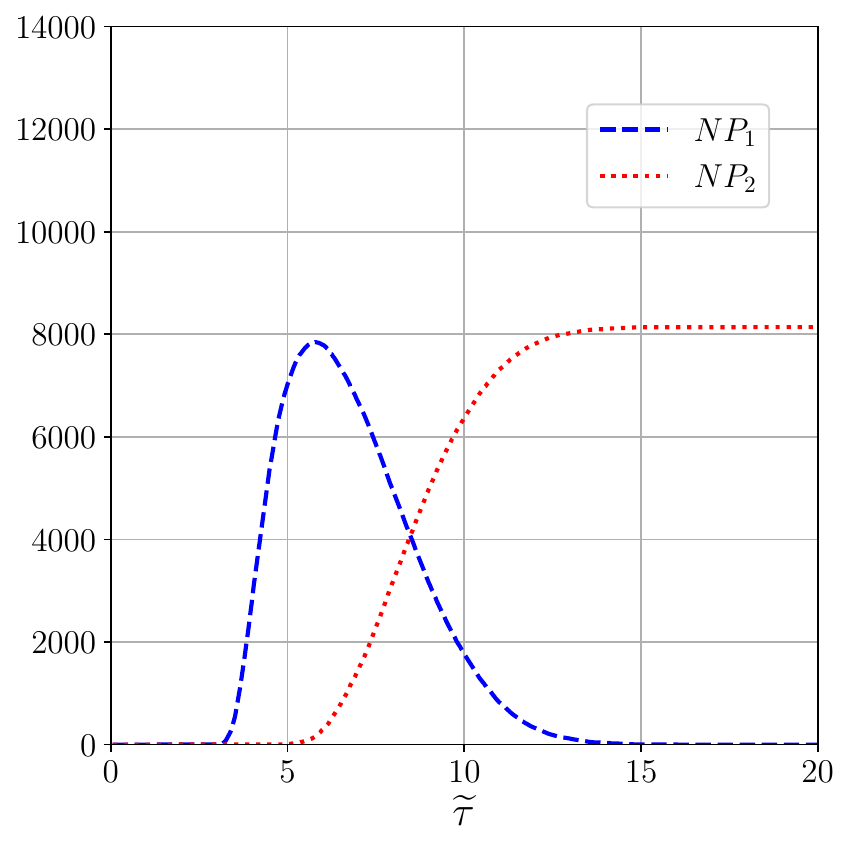}
	\end{minipage}
	\caption{
		The number count of the realisations out of $N = 10^{5}$ generated stochastic realisations for the two kinds of the window functions, $\widetilde{\mathrm{W}} (z) = \Theta (1 - z)$ and $\widetilde{\mathrm{W}} (z) = \exp \, ( - z^{2} / 2 )$ in the \textit{left} and \textit{right} panels respectively. 
        The parameters are fixed to be $\widetilde{k}_{\star} = 8$ and $\mathcal{P}_{0} = 1$. 
	}
	\label{fig:mf_hfil_a}
\end{figure}

To obtain these probabilities, the numerically generated $N = 10^{5}$ stochastic realisations (\textit{i.e.}~trajectories, part of which are shown in Figure~\ref{fig:rw_hfil_traj}) were scanned. 
That is, for each fixed time $\tau$, counting the number of realisations that (1) exceeds the threshold and (2) do not exceed the threshold but exceeded in the past, \textit{i.e.}~on a larger coarse-graining scale. 
The procedure was performed for the two kinds of window functions, and the recorded numbers of events $\mathrm{A}$ and $\mathrm{B}$ are shown in the left and right panels in Figure~\ref{fig:mf_hfil_a}. 
The parameters are fixed to be $\widetilde{k}_{\star} = 8$ and $\mathcal{P}_{0} = 1$ in both figure. 

In the left panel in Figure~\ref{fig:mf_hfil_a}, the Fourier-space Heaviside window function is used, where both the probabilities $P_{1}$ (dashed blue curve) and $P_{2}$ (dotted red curve) start to increase from zero almost simultaneously until around $\tau \simeq k_{\star}$, essentially confirming the result of Ref.~\cite{Kushwaha:2025zpz}, though note that the power spectrum there is slightly different from ours, Eq.~(\ref{eq:psG}). 
The reason that one realises $P_{1} (\tau) \approx P_{2} (\tau)$ in this regime is that, as was observed in Figure~\ref{fig:ncov_varH}, the ``up-crossing'' is driven mainly by the uncoloured noises, which is a peculiar feature when one uses the simplest window function, $\widetilde{\mathrm{W}} (z) = \Theta (1 - z)$. 
In other words, although the noises are coloured, the contribution from the uncoloured noise is dominant in the regime $\tau \lesssim k_{\star}$. 

In the regime $\tau \gtrsim k_{\star}$, on the other hand, the coloured noises push the stochastic trajectories back towards zero, so that the two probabilities start to behave differently. 
The dominance of the coloured noise in this regime can be observed from the asymptotic behaviours of the functions (\ref{eq:ncovG_iH}) in the equal-time limit. 
The function $I_{1} (\widetilde{\tau})$ that is directly related to the variance of $\delta (\tau)$ asymptotes to 
\begin{equation}
        I_{1} (\widetilde{\tau}) 
        \approx \frac{4}{\widetilde{\tau}^{6}} 
        \qty{ 
            (\widetilde{k}_{\star}^{4} + 9 \widetilde{k}_{\star}^{2} + 8) 
            \exp \bigg( - \frac{\widetilde{k}_{\star}^{2}}{2} \bigg) 
            + \widetilde{k}_{\star} 
            (\widetilde{k}_{\star}^{4} + 10 \widetilde{k}_{\star}^{2} + 15) 
            \cdot \sqrt{ \frac{\pi}{2} } 
            \qty[ 1 + \mathrm{erf} \bigg( \frac{\widetilde{k}_{\star}}{\sqrt{2}} \bigg) ] 
        } 
        \,\, , 
\end{equation}
while the other two to $I_{2} (\widetilde{\tau}) \approx 2 \exp \, [ - ( \widetilde{\tau} - \widetilde{k}_{\star} )^{2} / 2 ]$ and $I_{3} (\widetilde{\tau}) \approx ( \widetilde{\tau} / \Delta \widetilde{\tau} ) \exp \, [ - (\widetilde{\tau} - \widetilde{k}_{\star} )^{2} / 2 ]$, so $I_{1} (\widetilde{\tau})$ dominates over the others, again confirming Figure~\ref{fig:ncov_varH}. 
As a result, no realisation exceeds the threshold but part of them pierced it at some time in the past (\textit{i.e.}~on a larger scale), resulting in $P_{1} (\tau) \to 0$ and $P_{2} (\tau) \to \mathrm{const.}$ in the $\tau \to \infty$ limit. 

The right panel in Figure~\ref{fig:mf_hfil_a} shows the two probabilities, $P_{1} (\tau)$ and $P_{2} (\tau)$, when the Gaussian window function is used, so that the noises are completely correlated. 
In this case, as can be seen in Eqs.~(\ref{eq:ncovG_iG}), no uncorrelated noise exists and the coincidence $P_{1} (\tau) \approx P_{2} (\tau)$ is therefore not realised. 
Contrary to the case with $\widetilde{\mathrm{W}} (z) = \Theta (1 - z)$, the three functions (\ref{eq:ncovG_iG}) decreases in powers of $\tau$ in the regime $\tau \gg 1$. 
The difference is passed down to the mass function of PBHs as will be seen in Section~\ref{ssubsec:mf}. 

\subsubsection{Mass function of PBHs}
\label{ssubsec:mf}

\begin{figure}
	\centering
	\begin{minipage}[b]{0.495\linewidth}
		\centering
   \includegraphics[width=0.995\linewidth]{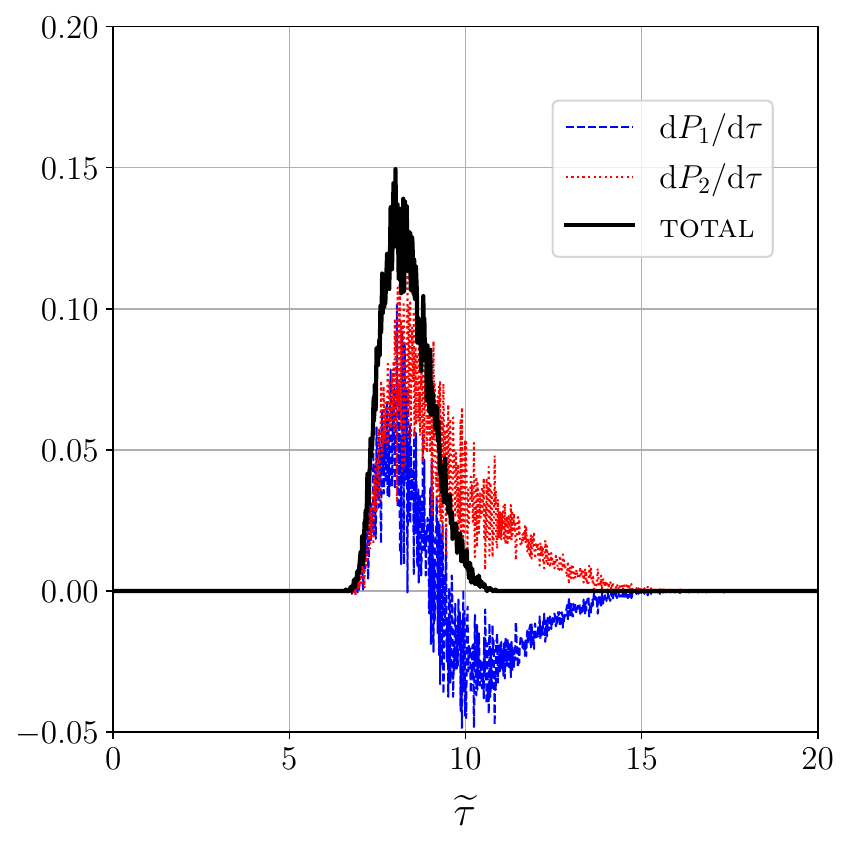}
	\end{minipage}
	\begin{minipage}[b]{0.495\linewidth}
		\centering
\includegraphics[width=0.995\linewidth]{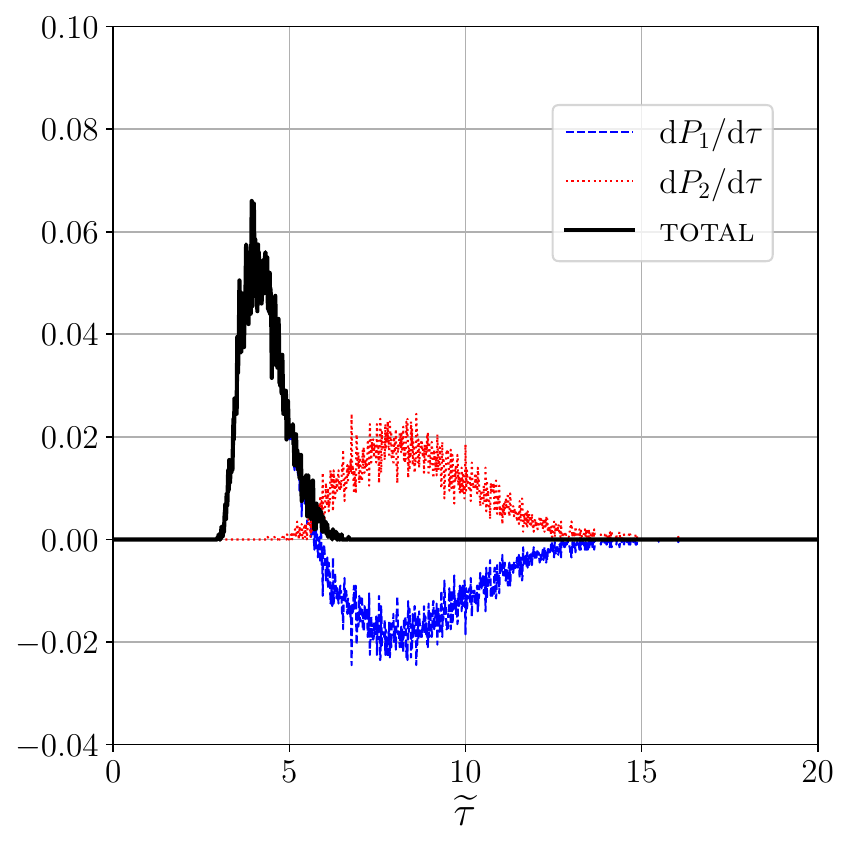}
	\end{minipage}
	\caption{
		The derivatives of the probability shown in Figure~\ref{fig:mf_hfil_a} for the two kinds of the window functions, $\widetilde{\mathrm{W}} (z) = \Theta (1 - z)$ and $\widetilde{\mathrm{W}} (z) = \exp \, ( - z^{2} / 2 )$ in the \textit{left} and \textit{right} panels respectively. 
        The parameters are fixed to be $\widetilde{k}_{\star} = 8$ and $\mathcal{P}_{0} = 1$. 
	}
	\label{fig:mf_hfil_b}
\end{figure}

The probability $P (\tau)$ is expected to count all the relevant structure that has the mass \textit{larger than} $M$. 
From $P (\tau)$, one defines the probability that a PBH of the mass within the infinitesimal range $(M - \dd M, \, M)$ is formed, 
\begin{equation}
	\dd P (\tau) 
	\equiv 
	P (\tau + \dd \tau) - P (\tau) 
	= \pdv{ P (\tau) }{ \tau } \, \dd \tau 
	\,\, . 
	\label{eq:pder}
\end{equation}
The mass function of PBHs is associated with the derivative in Eq.~(\ref{eq:pder}). 
To do so, $\tau$ must be expressed in terms of the mass $M$, but $\dd P (\tau) / \dd \tau$ and what is called the mass function differ only by a factor of order of some power of $\tau$, and thus the difference does not alter our consequences. 
Since the behaviour of the probabilities and their derivatives is of our primary interest, following Ref.~\cite{Kushwaha:2025zpz}, the derivative of $P (\tau)$, $\dd P (\tau) / \dd \tau$ in Eq.~(\ref{eq:pder}), is focussed on hereafter. 
    
From the probabilities discussed in Section~\ref{subsec:probs}, which are shown in Figure~\ref{fig:mf_hfil_a}, the mass function can be obtained also numerically. 
Figure~\ref{fig:mf_hfil_b} shows the derivatives of $P_{1} (\tau)$ and $P_{2} (\tau)$, as well as the sum $\dd P (\tau) / \dd \tau$. 
The parameters are again fixed for illustrative purposes, on which the following consequences do not depend. 
Several implications can be derived from those figures. 

First of all, as was already pointed out for $\widetilde{\mathrm{W}} (z) = \Theta (1 - z)$ in Ref.~\cite{Kushwaha:2025zpz}, not only $P_{1}$ but also $P_{2}$ must be accounted for to ensure the positivity of the mass function of PBHs. 
In addition to this, it can be confirmed that the necessity of both the probabilities is the case also when a smooth window function, in our study the Gaussian window function, is used. 
It is a direct consequence of the structured behaviour of the trajectory observed in Figure~\ref{fig:rw_hfil_traj}, which originates from the localised nature of the power spectrum, in addition to the evaluation at horizon crossing peculiar to PBHs, that give rise to the coloured noises. 
For the case with smooth window function, this nature is prominent only in the tail part, where the derivatives of the two probabilities compensate each other, as can be observed in the right panel in Figure~\ref{fig:mf_hfil_b}. 

Whilst the probability $P (\tau)$ consists of the two elements, the way of contribution of those probabilities $P_{1} (\tau)$ and $P_{2} (\tau)$ to the derivative of the total, $\dd P (\tau) / \dd \tau$, depends on which window function is used as can be seen in Figure~\ref{fig:mf_hfil_b}. 
In the left panel where $\widetilde{\mathrm{W}} (z) = \Theta (1 - z)$, the two kinds of the probabilities contribute to the total nearly equally. 
In other words, ignorance of $P_{1} (\tau)$ or $P_{2} (\tau)$ would non-negligibly affect the resultant abundance of PBHs. 
In the right panel where $\widetilde{\mathrm{W}} (z) = \exp \, (- z^{2} / 2)$, on the other hand, it can be observed that except for the deep tails $\dd P (\tau) / \dd \tau$ is determined solely by $\dd P_{1} (\tau) / \dd \tau$, corresponding to the fraction that the realisations exceed the threshold at a given mass scale. 
The difference between those two cases, between the non-smooth and smooth window functions, totally originates from the presence of the uncoloured noise. 
As was discussed in Section~\ref{subsec:probs}, in the former case, the contribution from the uncoloured noise dominantly drives the threshold-piercing that keeps $P_{1} (\tau) \approx P_{2} (\tau)$ in the regime $\tau \lesssim k_{\star}$, see the left panel in Figure~\ref{fig:mf_hfil_a}. 
This is no longer the case when a smooth window function is used. 
Indeed, there is no reason that realises a coincidence of the two probabilities since the noises are completely coloured. 
As a consequence, the two probabilities behave differently all the way from beginning to end. 
The stochastic trajectories driven by the coloured noises, in the absence of the uncoloured noise when $\widetilde{\mathrm{W}} (z) = \exp \, (- z^{2} / 2)$, pierces the threshold in a more structured way, although it is not of course fully deterministic, than the case with $\widetilde{\mathrm{W}} (z) = \Theta (1 - z)$. 
The mass function is therefore determined solely by $P_{1} (\tau)$, especially around the maximum. 
In summary, the relevant part (around the maximum) of the mass function of PBHs receives contributions from both events A and B when the Fourier-space Heaviside window function is used, whereas it is determined almost entirely by the event A for the smooth coarse-graining case.
This also implies that any cloud-in-cloud formation has negligible effect on the mass function around its maximum when a smooth window function is implemented. 

\section{Conclusion}
\label{sec:concl}

Primordial black holes, if exist, offer potential explanations for various cosmological conundrums, ranging from dark matter to early structure formation and gravitational-wave events. 
They also provide a probe of small-scale primordial quantum fluctuations, complementary to the large-scale inflationary fluctuations which have been stringently constrained by observations. 
However, in addition to the realisation of an exotic mechanism that gives rise to the formation of PBHs, 
there are many uncertainties that must be addressed to predict the mass distribution of PBHs. 
Amongst these, the choice of the window function is one of the causes that introduces a factor or order of uncertainties in the resultant mass distribution, one of the main focuses of the present article. 

The present article studied the formation probability of PBHs and the resultant mass distribution based on the excursion-set method. 
The stochastic noises are necessarily correlated due to the peculiar nature of PBHs, in which only the modes close to the horizon scale are relevant. 
As a result, even when the Fourier-space Heaviside window function is employed, the stochastic ``evolution’’ of the coarse-grained density contrast ceases to be Markovian, contrary to the standard excursion-set method. 
The absence of Markov property in general leads one to conduct numerical simulations, and in this article, the stochastic process in the presence of the coloured noise was numerically solved. 
The mass distribution under the two kinds of window functions, the Fourier-space Heaviside window function and the Gaussian window function, was then studied. 
Those two window functions enable us to derive the covariance matrix of the coloured noises analytically. 

Though any choice of the window function gives rise to correlated noises in the context of the formation of PBHs, there exists a contribution that mimics the uncorrelated noise when the Fourier-space Heaviside window function is used, while it is not the case when a smooth window function, such as the Gaussian window function, is implemented. 
The stochastic trajectories under the completely correlated noises are quite different from the situation where there exists a period in which the uncorrelated noises dominate the stochastic process, as can be observed in Figure~\ref{fig:rw_hfil_traj}. 
The resultant mass distribution is accordingly altered. 
It was in particular found that the total mass function is determined almost solely by one of the probabilities that corresponds to the threshold-exceeding fraction out of all the stochastic realisations. 
In other words, the derivative of the formation probability, $\mathrm{d} P (\tau) / \mathrm{d} \tau$ near its maximum, is dominated by the up-crossing fraction $N P_{1} (\tau)$, while the contribution from the fraction of trajectories that crossed the threshold at larger coarse-graining scales $N P_{2} (\tau)$ is negligible. 
This contrasts with the cases with the Fourier-space Heaviside window function, where both $N P_{1} (\tau)$ and $N P_{2} (\tau)$ contribute comparably. 
Our findings highlight that, the use of the excursion-set method in the context of the formation of PBHs, implementation of a smooth window function not only makes the noises completely correlated, but also leads to a physical impact on the resultant mass distribution of PBHs. 

Several future directions can follow the present study. 
First, amongst the three window functions that have been conventionally used in the literature, two of them were considered here in order to contrast the effect that comes from a smooth coarse-graining. 
It is thus straightforward to extend our analyses so that the relation between the mass function and the density contrast coarse-grained by the real-space Heaviside window function, the remaining conventional choice, is also included. 
Along with this line, converting the probabilities in terms of the mass of PBHs is another crucial direction to derive the phenomenological consequences based on the abundance of PBHs, which has been constrained by various observations. 

\acknowledgements

The authors are grateful to Chul-Moon Yoo for fruitful discussion. 
This work was supported by JSPS KAKENHI Grant Numbers JP24KJ1223 (DS) and JP24K22877 (KT). 

\bibliography{Bibliography}
\bibliographystyle{JHEP}

\end{document}